\documentstyle[12pt,aaspp]{article}

\begin{document}
\def\MSUN{\rm M_{\odot}}
\def\RSUN{\rm R_{\odot}}
\def\MSUNYR{\rm M_{\odot}\,yr^{-1}}
\def\MDOT{\dot{M}}

\newbox\grsign \setbox\grsign=\hbox{$>$} \newdimen\grdimen \grdimen=\ht\grsign
\newbox\simlessbox \newbox\simgreatbox
\setbox\simgreatbox=\hbox{\raise.5ex\hbox{$>$}\llap
     {\lower.5ex\hbox{$\sim$}}}\ht1=\grdimen\dp1=0pt
\setbox\simlessbox=\hbox{\raise.5ex\hbox{$<$}\llap
     {\lower.5ex\hbox{$\sim$}}}\ht2=\grdimen\dp2=0pt
\def\simgreat{\mathrel{\copy\simgreatbox}}
\def\simless{\mathrel{\copy\simlessbox}}

\title{Numerical simulations  
of mass outflows driven from accretion disks by  radiation  and 
magnetic forces.}

\vspace{1.cm}
\author{ Daniel Proga }
\vspace{1.cm}
\affil{JILA, University of Colorado, Boulder, CO 80309-0440, USA;
proga@colorado.edu}

\begin{abstract}

We study the two-dimensional, time-dependent magnetohydrodynamics (MHD)
of  radiation-driven winds  from luminous accretion disks initially threaded
by a purely axial magnetic field. The radiation force is mediated primarily 
by spectral lines and is calculated  using a generalized multidimensional 
formulation of the Sobolev approximation. We use ideal MHD to compute 
numerically the evolution of Keplerian disks, varying the magnetic field 
strengths and the luminosity of the disk, the central accreting object  
or both. 
We find  that the magnetic fields very quickly start deviating
from purely axial due to the magnetorotational 
instability. This leads to fast growth of the
toroidal magnetic field as field lines wind up due to the disk rotation.
As a result the toroidal field dominates over the poloidal field
above the disk and the gradient of the former drives a slow and dense disk 
outflow, which conserves specific angular momentum. 
Depending on the strength of the magnetic field relative to the 
system luminosity the disk wind can be radiation- or MHD driven.
The pure radiation-driven wind consists of a dense, slow outflow that is 
bounded on the polar side by a high-velocity  stream. The mass-loss rate is 
mostly due to the fast stream. As the magnetic field strength increases first 
the slow part of the flow is affected, namely it becomes  denser and slightly 
faster and begins to dominate the mass-loss rate. In very strong magnetic 
field or pure MHD cases, the wind consists of only a dense, slow outflow 
without the presence of the distinctive fast stream so typical to pure 
radiation-driven winds. Our simulations indicate that winds launched by 
the magnetic fields are likely to remain dominated by the fields downstream 
because of their relatively high densities. The radiation force due 
to lines may not be able  to change a dense MHD wind because the line force
strongly decreases with increasing density.

\end{abstract}


\section{Introduction}

Powerful mass outflows from accretion disks are observed in many 
astrophysical environments such as active galactic nuclei (AGN); 
many types of interacting binary stars, e.g., 
non-magnetic cataclysmic variables (nMCVs); and young stellar objects (YSOs).  
Magnetic fields, the radiation force and thermal expansion  
have been suggested as mechanisms that can drive disk winds.
These three mechanisms have been studied extensively 
using analytic as well as  numerical methods. 
As a result of these studies, theoretical models have been
developed that allow us to estimate under what physical conditions
each of these mechanisms is efficient in launching, accelerating
and collimating disk outflows. The notion that one universal physical
mechanism can explain all outflows
from accretion disks is very appealing.
However, theoretical studies and observational results indicate
that none of the three mechanisms alone is sufficient and
probably such a single mechanism does not exist.
Therefore it make sense to
consider a hybrid model in which more than one mechanism
is involved.

Magnetically driven winds from disks are the favored explanation for 
the outflows in many astrophysical environments.  Blandford \& Payne (1982) 
(see also Pelletier \& Pudritz 1992) showed that  the centrifugal force can  
drive a wind from  the disk if the poloidal component of the magnetic
field, ${\bf B_p}$ makes an angle of $> 30^o$ with respect to the normal to 
the disk surface. Generally, centrifugally-driven MHD  disk winds 
(magnetocentrifugal winds for short) require  the presence of a sufficiently
strong, large-scale, ordered magnetic field threading the disk 
with a poloidal component at least comparable to the toroidal magnetic
field, $|B_\phi/B_p| \simless 1$ (e.g., Cannizzo \& Pudritz 1988, 
Pelletier \& Pudritz 1992). Several groups have studied
numerically axisymmetric outflows using the Blandford \& Payne mechanism
(e.g., Ustyugova et al. 1995, 1999; Romanova et al. 1997; 
Ouyed \& Pudritz 1997a, 1997b, 1999; Krasnopolsky, Li \& Blandford 1999; 
Kato, Kudoh \& Shibata 2002).
An important feature of  magnetocentrifugal winds is that they require 
some assistance  to flow freely and steadily from the surface of the 
disk, to pass through a slow magnetosonic surface 
(e.g., Blandford \& Payne 1982). 
The numerical studies mentioned above do not resolve the vertical structure 
of the disk but treat it as a boundary surface through which mass 
is loaded on to the magnetic field lines at a specified rate.

Winds from disks can  driven by the magnetic pressure.
In particular, the toroidal magnetic field can quickly builds up
due to the differential rotation of the disk so that $|B_\phi/B_p| \gg 1$.
In such a case, the magnetic pressure of the toroidal field can give rise 
to a self-starting wind (e.g., Uchida \& Shibata 1985; 
Pudritz \& Norman 1986; Shibata \& Uchida 1986; Stone \& Norman 1994;
Contopoulos 1995; Kudoh \& Shibata 1997; Ouyed \& Pudritz 1997b).
To produce a steady outflow  driven by the magnetic pressure 
a steady supply of advected toroidal magnetic flux at the wind base 
is needed, otherwise the outflow is likely to be transient 
(e.g., K$\ddot{\rm o}$nigl 1993, Contopoulos 1995, Ouyed \& Pudritz 1997b). 
It is still not clear whether the differential rotation of the disk can 
produce such a supply of the toroidal magnetic flux to match 
the escape of magnetic flux in the wind and even 
if it does whether such a system will be stable  
(e.g., Contopoulos 1995, Ouyed \& Pudritz 1997b and references therein).

One of the reasons for favoring magnetic fields as an explanation
for mass outflows from accretion disks is the fact that magnetic fields 
are very likely crucial for the existence of all accretion disks.  
The magnetorotational instability (MRI)
(Balbus \& Hawley 1991; and earlier by Velikov 1959 and Chandrasekhar 1960)
has been shown to be  a very robust and universal mechanism to produce
turbulence and the transport of angular momentum in disks at all radii
(Balbus \& Hawley 1998). 
It is therefore likely that magnetic fields control mass accretion inside 
the disk and play a key role in producing a mass outflow from the disk. 
However, it has been demonstrated observationally and theoretically that
accretion disks are capable of losing mass also via  a radiation-driven wind, 
provided the disk luminousity in ultraviolet (UV) is high
enough.

Radiation-driven disk winds have been extensively modeled recently 
(e.g., Pereyra, Kallman \& Blondin 1997; Proga, Stone \& Drew 1998, 
hereafter PSD~98; Proga 1999; Proga, Stone \& Drew 1999, hereafter PSD~99; 
Feldmeier \& Shlosman 1999; Feldmeier, Shlosman \& Vitello 1999; 
Proga, Stone \& Kallman 2000; Pereyra, Kallman \& Blondin
2000; Proga \& Kallman 2002).
These studies showed that radiation pressure due to spectral lines 
can drive winds from luminous disks. This result
has been expected (e.g., Vitello \& Shlosman 1988). 
These studies, in particular those by PSD~98,  
also showed some unexpected 
results. For example, the flow is unsteady in cases where the disk
luminosity dominates the driving radiation field.
Despite the complex structure of the unsteady disk wind,
the time-averaged mass loss rate and terminal velocity scale
with luminosity, as do steady flows obtained where the
radiation is dominated by the central object. In 
the most favorable conditions (i.e., high
UV flux and low X-ray flux) the radiation force due to spectral lines
(the line force) can exceed the radiation
force due to electron scattering by a factor as high as $\sim 2000$
(e.g., Castor Abbott, \& Klein 1995, hereafter CAK; Abbott 1982; Gayley
1995). Thus systems with UV luminosity, $L_{UV}$ as low as  a few $10^{-4}$ 
of their Eddington limit, $L_{Edd}$, can produce a powerful high velocity wind.

Generally, one can argue 
that in all accretion disks, 
with $L_{UV} \simgreat$ a few $10^{-4} L_{Edd}$
mass outflows have been observed (Proga 2002). For example, accretion disks
around: massive black holes, white dwarfs (as in AGN and nMCVs with
$\Gamma_{UV} \ga 0.001$) and low mass young stellar objects 
(as in FU Ori stars with $\Gamma_{UV} \ga$ a few $\times$ 0.01) show 
powerful fast winds.
Systems that have too low  UV luminosities  to drive a wind 
include accretion disks around neutron stars and low mass black holes 
as in low mass X-ray binaries (LMXBs) and galactic black holes. 
These systems indeed 
do not show outflows similar to those observed in nMCVs, AGN and FU Ori. 
However, outflows, even in systems which appear to be luminous enough 
to produces radiation-driven disk winds,
cannot be fully explained by just line driving.

For example, Drew \& Proga (1999) applied results from pure line-driven 
(LD for short) disk wind models to nMCVs. In particular, 
they compared mass loss rates predicted by 
the  models with observational constraints. Drew \& Proga (1999) concluded 
that either mass accretion rates in high-state nMCVs are higher than presently 
thought  by a factor of 2-3 or that radiation pressure alone is not quite 
sufficient to drive the observed hypersonic flows. The difficulty in accounting
for the mass loss rate in a pure LD disk wind model for nMCVs is simply 
a reflection of the fact that the nMCV luminosities just barely satisfy 
the basic requirement, i.e., $L_{UV}\simless 7 \times 10^{-4} L_{Edd}$. 
Synthetic line profiles computed based on pure LD wind models confirmed 
Drew \& Proga's conclusion (Proga et al. 2002). 
Our study has been partially motivated by this conclusion
because if indeed radiation pressure alone does not suffice to drive 
the observed hypersonic flow then an obvious candidate to assist radiation 
pressure in these cases is MHD (e.g., Drew \& Proga 1999).

In this paper, we study how  magnetic fields can change disk winds driven 
by the line force for a given disk luminosity. 
We assume in our models that the transport 
of angular momentum in the disk is dominated by local disk viscosity, 
for instance due to the MRI in weakly magnetized disks 
(Balbus \& Hawley 1998). Here we  add  magnetic fields to the PSD~99 model 
and solve self-consistently the full set of ideal MHD equations.

The outline of this paper is as follows. We describe our calculations in 
Section 2. We present our results and discuss their perceived 
limitations in Section 3. The paper ends in Section~4,  with our conclusions.

\section{Method}

\subsection{Equations and Numerical Techniques}
To calculate the structure and evolution of a wind from a disk, we solve 
the equations of ideal MHD
\begin{equation}
   \frac{D\rho}{D t} + \rho {\bf \nabla \cdot  v} = 0,
\end{equation}
\begin{equation}
   \rho \frac{D{\bf v}}{Dt} = 
- {\bf \nabla P} - \rho {\bf \nabla} \Phi
+ \frac{1}{4\pi} {\bf (\nabla \times B) \times B}
 + \rho {\bf F}^{rad} 
\end{equation}
\begin{equation}\label{eqn:induction}
{\partial{\bf B}\over\partial t} = {\bf\nabla\times}({\bf v\times B}).
\end{equation}
Here the convective derivative $D/Dt$ is
equivalent to $\partial/\partial t + {\bf v\cdot\nabla}$.
The dependent quantities $\rho$, ${\bf v}$, and $P$ are gas mass
density, velocity, and scalar isotropic gas pressure,
respectively, and ${\bf B}$ is the magnetic field.
The gas in the wind is  isothermal with a sound speed $c_s$.
We calculate gravitational acceleration using the Newtonian potential $\Phi$.
The term ${\bf F}^{rad}$ in the equation of motion (2) is
the total radiation force per unit mass.
We solve these equations in spherical polar coordinates
$(r,\theta,\phi)$, assuming axial symmetry about the rotational axis
of the accretion disk ($\theta=0^\circ$)

The geometry and assumptions needed to compute the radiation field from 
the disk and central object are as in PSD~99 (see also PSD~98).
The disk is flat, Keplerian, geometrically-thin and optically-thick.  
We specify the radiation field of the disk by assuming that the local disk
intensity follows the radial profile of the so-called $\alpha$-disk 
(Shakura \& Sunyaev 1973), and therefore depends only on the mass accretion 
rate in the disk, $\dot{M}_a$, and the mass  and radius of the central object, 
$M_\ast$  and $r_\ast$.  In particular, the disk luminosity, 
$L_D \equiv GM_\ast \MDOT_a/2r_\ast$. In models where the central object  
radiates, we take into account the irradiation of the disk, assuming that 
the disk re-emits all absorbed energy locally and isotropically. We express 
the central object luminosity $L_\ast$ in units of the disk luminosity 
$L_\ast=x L_D$. 

Our numerical algorithm for evaluating the line force is described in PSD~99.  
Here we briefly describe the key elements of our calculations of the line
force. We use the  CAK force multiplier 
to calculate the line-driving force.
In this approximation, a general form for this force  
at a point defined by the position vector $\bf r$ is
\begin{equation}
{\bf F}^{rad,l}~({\bf{r}})=~\oint_{\Omega} M(t) 
\left(\hat{n} \frac{\sigma_e I({\bf r},\hat{n}) d\Omega}{c} \right),
\end{equation}
where $I$ is the frequency-integrated continuum intensity in the direction
defined by the unit vector $\hat{n}$, and $\Omega$ is the solid angle
subtended by the disk and central object at the point W. 
The term in brackets is the electron-scattering radiation force, where 
$\sigma_e$ is  the mass-scattering coefficient for free electrons,
and $M(t)$ is the force multiplier -- the numerical factor which
parameterizes by how much spectral lines increase the scattering
coefficient. In the Sobolev approximation, $M(t)$ is a function
of the optical depth parameter
\begin{equation}
t~=~\frac{\sigma_e \rho v_{th}}
{ \left| dv_l/dl \right|},
\end{equation}
where $v_{th}$ is the thermal velocity and 
$d v_l/dl$ is the velocity gradient along $\hat{n}$. 
The velocity gradient can be written as 
\begin{equation}
\frac{dv_l}{dl}=Q~\equiv~ \sum_{i,j}\frac{1}{2}\left(\frac{\delta v_i}{\delta r_j}
+\frac{\delta v_j}{\delta r_i}\right)n_in_j=\sum_{i,j}e_{ij}n_in_j,
\end{equation}
where $e_{ij}$ is the symmetric rate-of-strain tensor.
Expressions for the components of $e_{ij}$ in spherical polar coordinates
are given in Batchelor (1967).

We adopt the CAK  analytical expression
for the force multiplier as modified by Owocki, Castor \& Rybicki 
(1988, see also PSD98)
\begin{equation}
M(t)~=~k t^{-\alpha}~ 
\left[ \frac{(1+\tau_{max})^{(1-\alpha)}-1} {\tau_{max}^{(1-\alpha)}} \right]
\end{equation}
where $k$ is proportional to the total number of lines,
$\alpha$ is the ratio of optically thick to optically-thin lines,
$\tau_{max}=t\eta_{max}$, and $\eta_{max}$ is a parameter 
related to the opacity of the most optically thick lines.
The term in square brackets is the Owocki, Castor \& Rybicki correction
for the saturation of $M(t)$ as the wind becomes optically thin
even in the strongest lines, i.e., 
\begin{displaymath}
\lim_{\tau_{max} \rightarrow 0} M(t)~=~M_{max}~=~
k(1-\alpha)\eta_{max}^\alpha.
\end{displaymath}

We discretize the $r-\theta$ domain into zones in our calculation of
the wind structure. Our resolution in the $r$ and 
$\theta$ directions is sufficiently high  to ensure that the subsonic 
portion of the model outflow is sampled by at least a few grid points in 
both $r$ and $\theta$. This requirement and the nature of the problem combine 
to demand an increasingly fine mesh toward the disk plane: here the density 
declines dramatically with height, and, moreover, the velocity in the wind
increases rapidly.  Our numerical resolution consists of 100 zones in each of 
the $r$ and $\theta$ directions, with fixed zone size ratios, 
$dr_{k+1}/dr_{k}=d\theta_{l}/d\theta_{l+1} =1.05$.  

The boundary conditions for the hydrodynamic dependent quantities 
are specified as follows. At $\theta=0$, we apply an axis-of-symmetry 
boundary condition. For the outer radial boundary, we apply an outflow 
boundary condition.  For the inner radial boundary $r=r_{\ast}$ and for 
$\theta=90^o$, we apply reflecting boundary conditions.

Our simulations begin with a vertical, independent of $r$ magnetic field 
configuration and a Keplerian disk embedded in the rotating ambient medium 
of low density. The ambient medium is initially outflowing 
with the escape velocity  at $r=r_\ast$ in the direction
perpendicular to the disk midplane. Below we give the details of our initial
conditions and conditions in the first grid zone above the equatorial 
plane follows. 

We proceed with setting the initial conditions in the following way.
We start with adopting the rotational velocity
$v_\phi =\sqrt\frac{GM_\ast}{r \sin \theta}$ for $r \sin{\theta}> r_\ast$
and $v_\phi =0$ elsewhere. Thus the gas above the disk
rotates on cylinders with the disk Keplerian velocity whereas
the gas above the non-rotating central object has zero rotational 
velocity.

Our initial density profile is given by the condition of hydrostatic 
equilibrium in the latitudinal direction for a gas with
a given initial rotational velocity. To ensure an exact numerical
equilibrium initially, we first initialize the density
in the first grid zone above the equatorial 
plane, $\rho_0$. We assume $\rho_0$ is radius independent.
Then we integrate the latitudinal equation of motion from $\theta=90^\circ$
to $\theta=0^\circ$ to compute the pressure using the numerical difference
formula in our  code. Finally, we compute the density
from the isothermal equation of state.
To reduce the problems caused by very high 
Alfv${\acute{\rm e}}$nic velocities 
in regions of very low density
(i.e., to prevent the time step from being prohibitively small),
we set a lower limit to the density on 
the grid  as $\rho_{min}(r)=10^{-15} (r_o/r)^2~\rm 
g~cm^{-3}$ and enforce it at all times in all models.

For the initial poloidal velocity, we adopt 
$v_r = \sqrt{GM/r} \sin{\theta}$ and
$v_\theta = \sqrt{GM/r} \cos{\theta}$ for the region where 
the lower limit to the density
is  used (see above). This choice for the poloidal
velocity is motivated by practical concerns 
(i.e., again to prevent the time step from being prohibitively small) 
and is particularly useful
for models with no radiation force due to the central object.
For $x=0$, the gas above the disk collapses onto the disk and 
the time step becomes so small that the simulations practically stop.

To represent steady conditions in the photosphere at the base 
of the wind, during the evolution of each model we continued to apply 
the constraint that in the first zone above the equatorial plane 
the density is fixed at $\rho = \rho_0$ at all times.
During the evolution of our standard models,
$\rho_0$ was fixed at  $10^{-4}\rm g~cm^{-3}$.

There are several differences between our initial and boundary conditions 
for the hydrodynamic dependent quantities 
and those we used in previous work (PSD~98 and PSD~99).
The most important difference is in our treatment of the wind base.
In PSD~99,  to represent steady conditions in the photosphere at the base 
of the wind, during the evolution of each model we continued to apply 
the constraints that in the first zone above the equatorial plane 
the radial velocity $v_r=0$, the rotational velocity $v_\phi$ remains 
Keplerian, and the density is fixed at $\rho = \rho_0$ at all times.  
Physically, $\rho_0$ is analogous to the density in the photosphere 
of the disk at the base of the wind provided $\rho_0$ is relatively low. 
In PSD~99 and PSD~98, the interior of the disk itself was treated as 
negligibly thin and was 
excluded from the models (for a disk temperature of $10^{4}$~K at 
$r=2 r_\ast$, the disk scale height $H$ is $H/r_\ast \sim 10^{-3}$). 
In PSD~98 and PSD~99, the arbitrary value  for $\rho_0$ was fixed
typically at $10^{-9} \rm g~cm^{-3} $.  As discussed in 
PSD~98,  the gross properties of  LD winds are unaffected by the 
value of $\rho_0$ provided it is large enough that the acceleration of the 
wind up to the sonic point is resolved with at least a few grid points.
This technique, when applied to calculations of spherically symmetric
LD winds from stars, produces a solution that relaxes to the
appropriate CAK solution within a few dynamical crossing times.
However, applying these constraints in MHD simulations can cause
very strong  evolution of  the gas close to the equator and 
consequently the evolution of the mass outflow.
We have performed many tests and found that,
when the above constraints are used, the disk gets
disrupted very quickly (within a couple orbits at $r=r_\ast$) by  MRI,
making it impossible to study disk winds.
Inclusion of magnetic fields in the model puts
new constraints on  $\rho_0$.
For strong magnetic fields and low $\rho_0$ (i.e., a small value of 
the plasma parameter $\beta\equiv 8\pi P/ B^2$ on the equatorial plane), 
the MRI or magnetic braking 
can very quickly reduce the density near the disk midplane.
In particular, we observe a dramatic drop in the density, of 
several orders of magnitude,
between the first and second zone above the equatorial plane.
This change in the density profile
makes impossible to study outflows from a `steady state' disk.
By choosing a large $\rho_0$ for a given magnetic field, we can reduce 
the dynamical importance of the magnetic field in
the region close to the midplane so that the base of the wind can remain 
in a steady state for a long time.
For these reasons, our value of $\rho_0$ is 
$\rho_0=10^{-4}\rm g~cm^{-3}$ and during the evolution of our standard models
we allow all the dependent quantities (including $\rho$, $v_r$, and 
$v_\phi$) to float everywhere on the grid.
For the adopted value of $\rho_0$, the plasma parameter on the 
equator, $\beta_0\equiv 8\pi P/ B^2=8\pi c^2_s \rho_0/ B^2$ is 
very high for all standard models 
(i.e., $ 2\times 10^5 \leq \beta_0 \leq 2 \times 10^{11}$).
In Section~3.3, we discuss 
how our results depend on the conditions
along the equatorial plane, in particular 
what difference  it makes if 
$\rho_0\leq10^{-9}\rm g~cm^{-3}$, 
and $\rho$, $v_r$,  and $v_\phi$ are set as in PSD~99.

We are left with describing our initial magnetic configuration and the 
boundary conditions for the magnetic field.
We assume a force-free configuration (Lorentz force
$\bf{ (\nabla \times B) \times B}=0$) by simply setting
$\bf (\nabla \times B)=0$. We consider one straightforward
initial magnetic configuration satisfying these constraints:
a uniform vertical field configuration defined by the
magnetic potential ${\bf A} = (A_r=0, A_\theta=0, A_\phi= A r \sin\theta)$.
We scale the magnitude of the magnetic field
using a parameter, $\beta'_w \equiv 32 \pi^2 c_s^2 \rho_w/B^2$ 
defined for a fiducial wind density of
$\rho_w= 10^{-15}\rm g~cm^{-3}$:
\begin{equation}
A =2~\pi~\sqrt{(2 c^2_s \rho_w/\beta'_w)}.
\end{equation}

The boundary conditions for the magnetic field are: at
$\theta=0$, we apply an axis-of-symmetry boundary condition;  for
the outer radial boundary, we apply an outflow boundary 
condition.  For the inner radial boundary $r=r_{\ast}$
we apply reflecting boundary conditions
while for $\theta=90^o$ we apply an equatorial-symmetry boundary condition
(Stone \& Norman 1992b).
In our standard models we allow all three components of
the magnetic field to float. 

In reality, gas near the disk midplane (inside the disk) is turbulent because 
of the MRI but in near hydrostatic equilibrium. We would like to stress that, 
although we include the region very close to the midplane in
our standard models, we do not claim that we model the disk 
interior. To do the latter we would need to add physical processes such 
as magnetic field dissipation and radiative transfer appropriate to
optically thick disk gas, and to solve the equation of energy. 
We would also need to resolve better the disk so we could capture, 
for example, the fastest growing modes of the MRI. 
The most unstable wavelength of the MRI, $\lambda \sim 2 \pi v_A/\Omega$,
increases with height in the disk for a given angular velocity, 
$\Omega$, because the Alfv${\acute{\rm e}}$n speed increases with height. 
In general,  near the equator, we deal with a magnetized stratified disk
where magnetic field is generated either by the MRI or magnetic braking.
It is computationally prohibitive to resolve adequately a thin disk in global
calculations even in two dimensions. Nevertheless, as we will discuss 
in section~3.3, our simulations are consistent with  high resolution local 
simulations of magnetized stratified disks (e.g., Stone et al. 1996; 
Miller and Stone 2000). 

To solve eqs. (1)-(3), we use the ZEUS-2D code described by 
Stone \& Norman (1992a, 1992b).

\subsection{Model Parameters}

As in PSD~99 and Proga (2000), we calculate disk winds with model parameters 
suitable for a typical nMCV (see Table~1 in PSD~99, Table~1 in Proga 2000 
and our Table~1). We vary the disk and central object luminosity and 
the strength of magnetic field (i.e., $\beta'_w$). We hold all other 
parameters fixed: $M_\ast=0.6~\MSUN$, $r_\ast=8.7\times10^8~\rm{cm}$,
$c_s=14~\rm{km~s^{-1}}$, $k=0.2$, $\alpha=0.6$ and $M_{max}=4400$
(see Table~1 in PSD~98). 
Nevertheless we can use our results to predict the wind 
properties for other parameters and  systems -- such as AGN and YSOs -- 
by applying the dimensionless form of the hydrodynamic equations and 
the scaling formulae as discussed in PSD~98 and Proga (1999).
We note that by adopting the uniform vertical field configuration,
our model has  two free parameters more than the pure LD
disk wind model of PSD~99: the plasma parameter $\beta'_w$ and
the initial toroidal magnetic field, $B_\phi$. In this paper, we
hold $B_\phi=0$ but vary $\beta'_w$. 

Summarizing, our model has three crucial independent parameters: 
the Eddington factor corresponding to the disk luminosity
$\Gamma_D\equiv L_D/L_{Edd}=\frac{\sigma_e \MDOT_a}{8\pi c r_\ast }$;
the central
object luminosity expressed in terms of the disk luminosity,
$x=L_D/L_\ast$, and the strength of the magnetic
field expressed in terms of the plasma parameter
for a fiducial density of $\rho_w= 10^{-15}\rm g~cm^{-3}$, $\beta'_w$.
As in PSD~99,
we assume that all the radiation is emitted in the ultraviolet and does
not evolve.
The model also has other parameters that can be 
calculated self-consistently, in principle, 
for a given set of $\Gamma_D$, $x$, $\beta'_w$ 
and the spectral distribution of the radiation field.
These parameters are: the sound speed $c_s$, and 
the parameters of  
the CAK force multiplier, $k$, $\alpha$ and $M_{max}$.
For the pure LD case, the wind solution
does not depend on the value of the sound speed (see CAK for 
LD stellar winds and  Proga 1999 for LD disk winds). 
The basic requirement for the strength of the disk radiation
to drive a wind is: 
\begin{equation}
\Gamma_D [1+ M_{max}] > 1
\end{equation}
(e.g., PSD~98; Proga 1999; 
see also Proga \& Kallman 2002 and Proga 2002 for a discussion
of the generalized version of this requirement).
Therefore,  $k$ and $\alpha$ do not matter 
as much as the value of $M_{max}$.

\section{Results}

Pure LD winds from a disk fall into two categories: 
1) intrinsically unsteady with large fluctuations in density and velocity, 
and 2) steady with smooth density and velocity  (PSD~98 and PSD~99). 
The type of flow is set by the geometry of the radiation field, 
parametrized by $x$: if the radiation field is dominated by the disk ($x<1$) 
then the flow is unsteady, and if the radiation is dominated by the central 
object ($x\simgreat 1$) then the flow is steady. The geometry of the radiation 
field also determines the geometry of the flow: the wind becomes more polar 
as $x$ decreases. However, the mass-loss rate and terminal velocity 
are insensitive to geometry and depend more on the total system luminosity, 
$L_D+L_\ast$. Regardless of the type of  flow,  pure LD
winds consist of a dense, slow outflow that is bounded on the polar side by a
high-velocity  stream. The mass-loss rate is mostly due to the fast stream.

In Proga (2000), we recalculated some of the PSD~99 models to check 
how inclusion of the magnetocentrifugal force, corresponding
to purely poloidal $\bf{B}$, will change LD disk winds. 
We found that flows which conserve  specific angular velocity have a larger 
mass loss rate  than their counterparts with purely LD flow, 
which conserve specific angular momentum. The difference in the mass loss rate 
between  winds conserving specific angular momentum and those conserving
angular velocity can be several orders of magnitude for low disk luminosities 
but vanishes for high  disk luminosities. Winds which conserve angular 
velocity have much higher velocities than angular momentum conserving winds.
In Proga (2000), we  also found that fixing the wind geometry stabilizes winds 
which are unsteady when the geometry is derived self-consistently.
Additionally, as expected, the inclination angle $i$ 
between the poloidal velocity 
and the normal to the disk midplane is important. 
Non-zero inclination angles allow the magnetocentrifugal force to increase
the mass loss rate for low luminosities, and increase the wind velocity
for all luminosities. 

In this paper, we also recalculate some of the PSD~99 flows,  but this time 
we check, by solving the  full set of ideal MHD equations and allowing 
$B_\phi \neq 0$,  how {\it self-consistent} inclusion of magnetic fields to 
the PSD~99 model will change LD disk winds. We summarize the properties of 
PSD~99 models and our new simulations in Table~1. Columns (2) to (4) give 
the input parameters that we varied: the mass accretion rate $\MDOT_a$, 
the relative luminosity of the central object, $x$, and the plasma parameter
$\beta'_w$, respectively. Column (5) lists the final time at which we stopped 
each simulation (all times here are in units of 
$\tau=\sqrt{r^3_\ast/G M_\ast}=~2.88~ \rm sec$). Columns (6) to (8) give
some the gross properties of the disk wind: the wind mass loss rate, 
$\MDOT_w$, the wind velocity at the outer radial boundary, $v_r(10 r_\ast)$, 
and the wind half-opening angle, $\omega$, respectively.
We measure $\omega$ from the equator to the upper envelope of the wind.
Table~1 also  contains comments regarding some runs [column (9)]
and explains  our convention of  labeling our runs.

\subsection{Outflow from a luminous magnetized accretion disk}

In this section we describe the properties
and behavior of our model MHD-LD C0D in which 
$\MDOT_a~=~10^{-8}~\MSUNYR$, $x=0$,  $\beta'_w=2\times10^{-2}$. 
This model is a rerun of the fiducial `$x=0$' unsteady model
discussed in detail in PSD~98 and PSD~99.

Figure~1 presents a sequence of maps showing density, velocity field and
toroidal magnetic field (left, middle and  right panels) from model C0D, 
plotted in the $r,z$ plane.  The length of the arrows in the middle panels is
proportional to $(v_r^2 + v_\theta^2)^{1/2}$. To show better the evolution
of the wind with lower velocities we use the maximum lenght of the arrows 
in regions of high velocity, i.e., $(v_r^2 + v_\theta^2)^{1/2}\ge 
200~\rm km~s^{-1}$.  We also suppress velocity vectors in regions
of low density (i.e., $\rho$ less than $10^{-15}~\rm g~cm^{-3}$).
The pattern of the direction of the arrows is an 
indication of the shape of the instantaneous streamlines.  The solid lines in 
the middle panels mark the location where the poloidal Alfv${\acute{\rm e}}$n 
speed, $v_{Ap}\equiv B_p/\sqrt{4\pi \rho}$, equals the poloidal fluid speed, 
$v_p$ (i.e. the Alfv${\acute{\rm e}}$n surface). As in the pure LD case, after 
$\sim 10$ time units disk material fills the grid for $\theta \simgreat 30^o$ 
and remains in that region for the rest of the run. In the early phases of 
the evolution the MHD-LD disk wind resembles its pure LD counterpart except 
that it is steadier. However, in the late phases the flow undergoes a dramatic 
change not seen in the LD case: slow and dense material rises from 
the disk at large radii. 
In the slow wind, the density increases whereas the velocity
decreases with time until the wind settles to a time-averaged 
steady state. These changes in the slow wind occur on relatively 
a long time scale and are caused by the evolution of the magnetic field. 
In particular, $B_\phi$ is generated near the base of the wind
as indicated by the rise of the $B_\phi$ contours in Figure~1 
(the strength of $B_\phi$ decreases with $z$). 
The gradient of $B^2_\phi$ drives the slow wind 
from the disk at large radii. 
At the end of the simulation, the LD wind 
is replaced by a wind driven primarily by the magnetic pressure.

Figure~2  presents the density and poloidal velocity field in the wind at 
the end of our simulations at 680~$\tau$ (top left and middle panels). 
The dashed and solid line in the top right panel of Figure~2 shows 
the contours of the angular velocity $\Omega$ and specific angular momentum 
of fluid, $L\equiv r \sin{\theta} v_\phi$, respectively.
The bottom left, middle and right panels of Figure~2 show
the contours of the $\beta$ plasma parameter, 
the poloidal magnetic field and the contours of the toroidal magnetic
field, $B_\phi$, respectively. The solid lines in the bottom middle
panel show the location where the strength of the poloidal magnetic 
field equals  the toroidal magnetic field, $\mid B_p \mid =\mid B_\phi \mid$.

Comparing our model C0D with its pure LD counterpart, we find that model C0D 
has a relatively smooth density distribution and  is relatively steady. 
For example, the pure LD outflow is intrinsically unsteady and characterized 
by large amplitude velocity and density  fluctuations.  Infall as well as 
outflow from a disk can occur in different regions of the wind at 
the same time. However, in model ~C0D  the poloidal velocity
is mostly organized and there is no inflow onto the disk in the wind
domain. Model~C0D has not
reached a steady state even after $400~\tau$. However, we observe that 
this model is  close to reaching such a state; for example, the mass flux 
density settles to some time-averaged maximum.

As in the pure LD case, we find that the wind consists of 
a dense, slow outflow that is bounded on the polar side by a
high-velocity  stream. However, there is an important difference, namely that
the dense, slow outflow is significantly denser and somewhat faster
in our model C0D than the pure LD case. 
Finally, as one could have expected for an MHD wind, 
the gas pressure dominates over the magnetic
pressure in the region very close to the mid-plane ($\beta \gg 1$) 
whereas the opposite is true in the wind domain ($\beta \ll 1$,
see bottom left panel).

The above differences in the two wind solutions appear to result from
the Lorentz force due to the gradient of the  toroidal magnetic field. 
Our simulations start with zero $B_\phi$ but this situation changes very 
quickly as toroidal field is generated by  rotation. We note that 
${\bf \nabla} B^2_\phi$  is higher than the line force in the wind domain 
(expect for the fast stream) by a couple of orders of magnitude. Comparing 
the contours of $B_\phi$ (bottom right panel of Fig. 2) and the poloidal 
velocity field (top middle panel), we find that the velocity field is normal 
to the  $B_\phi$ contours (i.e., $v_p$ is parallel to the $B^2_\phi$ gradient)
in the slow wind (lower right-hand corner of each panel). In contrast 
to the slow wind, in the fast stream where the gas is driven by 
the line force and the magnetic pressure is unimportant, $v_p$ is tangent 
to the $B_\phi$ contours.

The quantities presented in Figures~1 and 2 show that the outflow in model~C0D
is not a magnetocentrifugal wind. There are several diagnostics of 
magnetocentrifugal driving. For example, in the acceleration zone
of a steady state magnetocentrifugal wind $B_p \simgreat B_\phi$. 
However in model C0D, the location where $\mid B_p \mid =\mid B_\phi \mid$ 
is near the disk and 
$B_p < B_\phi$ in most of the acceleration zone 
(see lower middle panel in Figure~2).
Additionally, in a magnetocentrifugal wind the total angular
momentum per unit mass, 
$ l \equiv v_\phi r \sin{\theta}- r \sin{\theta} B_\phi B_p/(4\pi \rho v_p)$, 
the second term due to the twisted magnetic fields is comparable
to the first term.
Subsequently, the wind is corotating with the underlying disk
up to approximately the Alfv${\acute{\rm e}}$n point. In model ~C0D, 
the Alfv${\acute{\rm e}}$nic 
surface is very close to the disk (see top middle panel) 
and the total specific angular momentum is mostly due to the fluid
even near the wind base. Thus the wind is corotating with the disk only
over a relatively small length scale. The top right panel
of Figure~2 is helpful to distinguish between a magnetocentrifugal
wind and a fluid angular momentum conserving wind. In the former,
the angular velocity is conserved along the streamlines below 
the Alfv${\acute{\rm e}}$n point while in the latter, $L$ is conserved along 
the streamlines. Comparing the top right panel 
and the top middle panel of Figure~2, we  clearly
see that the contours of $L$,  not of $\Omega$, are aligned with
the streamlines represented by 
the arrows of the poloidal velocity field.

Next we consider the angular dependence of the flow at large radii.
Figure~3 shows the angular dependence of 
density, radial velocity, mass flux density, and accumulated mass loss rate 
at $r = r_o = 10 r_{\ast}$ at the end of the simulation of model~C0D.
The accumulated mass loss rate is given by:
\begin{equation}
d\dot{m}(\theta) =
4 \pi r_o^2 \int_{0^o}^\theta \rho v_r \sin \theta d\theta.
\end{equation}
The gas density is a very strong function of angle for $\theta$ between
$90^o$ and $25^o$.  Between the disk mid-plane at $\theta = 90^o$ and 
$\theta \sim 85^o$, $\rho$ drops by $\sim 7$ orders of magnitude, as  expected
for a density profile determined by hydrostatic equilibrium. 
For $25^o \simless \theta \simless 85^o$, the wind domain, $\rho$ varies 
between $10^{-16}$ and $10^{-11}~\rm g~cm^{-3}$.  For $\theta \simless 25^o$, 
density  again decreases rapidly, but this time to so low a value that 
it becomes necessary to replace it by the numerical lower limit $\rho_{min}$.
The region with $\rho \le \rho_{min}$ is not relevant to our analysis as it 
has no effect on the disk flow.  The radial velocity at $10r_\ast$
increases gradually from  zero at the equator to 
 $\sim 100~\rm km~s^{-1}$ at $\theta\approx60^o$, then it drops to nearly zero
at $\theta\approx60^o$.
Over the angular range  $65^o > \theta >
25^o$, $v_r$ increases from $\leq 0$ up to $1200~\rm km~s^{-1}$.

The cumulative mass loss rate is negligible for $\theta \simless 35^o$
because of the very low prevailing gas density.  Beginning at $\theta 
\simgreat 35^o$, $d\dot{m}$ increases to $\sim 3\times 10^{13}~\rm g~s^{-1}$ 
at $\theta \approx 55^o$. This increase of  $d\dot{m}$
is due to the fast stream. Then, in the slow dense outflow, 
the cumulative mass loss rate increases to $\sim 7\times 10^{14}~\rm
g~s^{-1}$ at $\theta \approx 86^\circ$. For even higher $\theta$, 
in the region close to 
the disk plane, where the gas density starts to rise very sharply and where 
the motion is subsonic and  typically more complex, 
the cumulative mass loss rate is 
subject to enormous fluctuations (some of which may even be negative).
In the example shown in Figure~3, the total mass loss rate through the outer 
boundary, $\MDOT_{tot}=$ $d\dot{m}(90^o)$ reaches $\sim  10^{19}~\rm 
g~s^{-1}$!  This value of $\MDOT_{tot}=$ 
is most certainly dominated by the contribution from 
the slow-moving region very close to the disk mid-plane -- a contribution 
that is very markedly time-dependent. We ignore the value of $\MDOT_{tot}$
as it is related more to subsonic oscillations of the 'disk' rather
than to the supersonic wind we model. In the remaining part of the
paper, for the wind mass loss rate we use the value of 
the cumulative mass loss rate at $\theta=82^\circ$.

We note  that the increase of the mass loss rate
in the slow wind is totally due to magnetic fields and  there
is no enhancement of the line force in this region. In fact,
the opposite is true -- the line force in the dense slow region in model~C0D
is significantly reduced by the action of the magnetic fields.
The latter increases the density of the outflow which in turn
reduces the line force. The slight increase in the velocity 
and its gradient in the slow wind is far from compensating
the reduction of the line force due to 
the increase of the density.

We conclude that magnetic fields can change qualitatively
and quantitatively a radiation-driven disk wind. In particular,
the magnetic pressure can dominate the driving of the  wind and reduce
the role of the line force. In model ~C0D, we find
that a disk loses mass via a LD wind 
in the inner part of the wind (the fast stream) and via a MHD-driven  wind
in the outer part of the wind (the dense slow wind).
We expect  that increasing the relative strength of the magnetic pressure 
to the radiation pressure 
(e.g., by reducing $\beta'_w$ and therefore $\beta_0$)
should lead to
the entire wind domain being driven by the magnetic force.
Conversely, decreasing the magnetic pressure should lead to 
the wind being driven by the line force.  Therefore we consider next a limited 
parameter survey to check  whether our expectation is correct
and to see how   the wind solution
changes  quantitatively with the strength of the radiation and magnetic fields.

\subsection{Parameter survey}

We consider only the parameter space of our models that will 
define the major trends in disk wind behavior.  
We focus on a survey of how the mass loss rate, outflow velocity and geometry
change with disk luminosity, relative central object luminosity and
strength of magnetic field. In Section~3.3 we will consider
how our results depend on our treatment of the conditions in the first
grid zone above the equatorial plane.

In Figure~4 we show (a) the wind mass loss rate, $\MDOT_w$, as a 
function of the total Eddington factor, $(1+x)\Gamma_D$ 
(note that $(1+x)\Gamma_D$ is proportional $\MDOT_a$ 
for a given $x$) for various $\beta'_w$ and (b) $\MDOT_w$, as a 
function of $\beta'_w$,  for  various $(1+x)\Gamma_D$.  
The top panel of Figure~4 also shows 
$\MDOT_w$ as a function of $(1+x)\Gamma_D$ 
as predicted by the pure LD wind model with and without taking into account
the fact that the force multiplier has a maximum value.
The thick  solid line corresponds to the prediction where $M(t)$ can
be arbitrarily high whereas the thin solid line corresponds
to the prediction where $M(t)$ reaches maximum at $M_{max}=4400$
(see Proga 1999).
In the top panel of  Figure~4, it can be seen that  
$\MDOT_w$ is a very strong function of $(1+x)\Gamma_D$ for $\beta'_w=\infty$.  
As shown in PSD~98 and Proga~(1999), two dimensional models
of LD disk winds predict mass loss rates (as well as velocities)
very similar to those predicted by the original CAK formulae
when the stellar Eddington factor is replaced with 
the the total Eddington factor and the formulae
are corrected for the fact that for small Eddington factors
the force multiplier reaches its maximum and consequently
the radiation force cannot exceed gravity.
Note that the drop in $\MDOT_w$ occurs 
for $(1+x)\Gamma_D\approx 0.0002=1/M_{max}$.

Motivated by the conclusion from section~3.1, we have
performed a few simulations using different values of $\beta'_w$.
We would like to estimate a  $\beta'_w$ range for which
the line force dominates and magnetic fields play a small
role,  and a range for which the line force is unimportant
and the wind is totally controlled by magnetic fields.

We here focus on simulations for a fiducial `$x=0$' model with 
$\MDOT_a=10^{-8}~\MSUNYR$
and a fiducial `$x=1$' steady state model with $\MDOT_a=\pi \times 
10^{-8}~\MSUNYR$ and various $\beta'_w$.
For both series of simulations 
(i.e., series `C0' corresponding to runs labeled 
C0A, C0B, C0D, C0E, and C0F
and series `D1' corresponding to runs labeled 
D1A, D1B, D1D, D1E, D1F, and D1G, see Table~1),
we find that indeed for sufficiently high $\beta'_w$ the wind mass loss 
rate and characteristic velocity 
are very similar for those in the pure LD counterparts. For example,
we note that for models of series~D1,
the symbols corresponding to $\MDOT_w$ with $\beta'_w=2, 2\times10^{-2}$,
and $2\times10^{-3}$ overlap with the the symbol corresponding
to the pure LD model D1A (see also the bottom panel of Figure~4). However,
there is a qualitative difference between the MHD-LD winds
and the LD wind even for high $\beta'_w$ cases: the MHD-LD winds
are less unsteady than the LD winds. For high $\beta'_w$, the Lorentz
force can be too weak to drive a wind but if $\beta\ll 1$ 
in the wind (as in the cases we explored), the gradient of the
magnetic pressure can reduce the density fluctuation
near the wind base and in the wind.

The bottom panel of Figure~4 shows more clearly the point we made above
that for high $\beta'_w$, $\MDOT_w$ is as for 
pure LD cases and does not change with $\beta'_w$. 
However, starting 
from a certain value of $\beta'_w$, $\MDOT_w$ becomes a strong
function of $\beta'_w$. The value of $\beta'_w$ at which
$\MDOT_w$ starts to increase with decreasing $\beta'_w$ 
is higher for the lower luminosity system (the solid line) 
than for the higher luminosity system (the dashed line).
Generally, 
the strong $\MDOT_w$ dependence on $(1+x)\Gamma_D$ almost disappears
when strong magnetic fields are added to the model and this
is clear in the top panel, where $\MDOT_w$ becomes insensitive to
$\Gamma_D$ as $\beta'_w$ decreases.

As Table~1 shows, the wind geometry is also sensitive to $\beta'_w$:
the half-opening angle of the wind $\omega$, increases with decreasing
$\beta'_w$ so
the wind becomes more polar as $\beta'_w$ decreases. We note that
even the `$x=1$' models become polar for $\beta'_w,2\times10^{-4}$
(see models D1F and D1G).

Simulations for fixed $\beta'_w$ and $x$, but with varying the system 
luminosity,  help us to check whether the magnetic fields
can drive a wind for disk luminosities too low to the drive a LD wind.
These simulations also help us to check whether the line force
can `regain' control over the wind  if the system luminosity increases.

For the system luminosity  of model B1A, the line force 
is too weak to produce a supersonic outflow (e.g., PSD~99). However, 
in model B1F with the same radiation field as in model B1A, there is a strong 
robust disk outflow. 
In the models B1F and D1F with the same low $\beta'_w$
but different system luminosities, the wind is mostly driven 
by  MHD. However in model E1F, with the system luminosity higher
than in model D1F by a factor of 10 
the wind is LD and very similar to that with zero magnetic field 
(compare model E1F with model E1A).

We conclude that MHD driving is robust and does not require the line
force, e.g., 
the MHD driving does not require the line force to launch a wind from 
the disk. We find a negative feedback between 
the magnetic field and the line force, i.e., the higher the magnetic field 
the lower the line force and vice versa. 
We also conclude that although  MHD driving can produce
a strong wind it does so by driving a relatively dense wind with relatively 
low velocities. The latter do not seem to depend on the escape velocity
from the radius from which the wind is launched. 
We note that $v_r$ of the slow wind depends very weakly on $\theta$
at $r_o$. On the other hand, the line
force drives a relatively fast wind with a velocity sensitive
to the launching radius and consequently to $\theta$ (PSD~98).

\subsection{Dependence of wind evolution and properties on the treatment
of the wind base}

Our model for mass outflows from accretion disks is a hybrid
of an LD model  and an MHD-driven flow.  We have extensively tested our
LD disk wind model and applied it to several systems (see references in
Section~1). In this section we present a brief review of  our
test runs designed to check the MHD part of our model.
In particular, we have performed a few tests aimed at reproducing
qualitatively results already published on MHD disk winds.
Additionally, we have explored a parameter region of our model
($0.1\simless \beta_0 \simless 10$) for which we expect to resolve
the fastest growing modes of the MRI.

There have been many numerical studies of MHD disk winds 
(our list of references  in the introduction is far from complete).  
For simplicity of our presentation we will reference
in more detail to the work of only two groups: Stone \& Norman (1994)
and Ouyed \& Pudritz (1997a; 1997b; 1999). The former
included the disk as well as the wind in their simulations whereas
the latter included only the wind and treated the disk as the lower
boundary of the wind. Both approaches have been commonly 
adopted in the literature.

Clearly, inclusion of the disk structure in calculations of disk outflow
is highly desirable. However, the physics of the disk is very complex
and its proper modeling is very demanding. On the other hand, one
would hope that it is possible to capture the key elements of a disk
wind without modeling the disk interior. 
Work by Ouyed \& Pudritz (1997a; 1997b; 1999) is an example
of studies of  magnetocentrifugal disk winds while PSD~98, PSD~99 
is an example
of studies of LD disk winds where the disk interior was not included.
However there is an important difference between modeling
a magnetocentrifugal disk wind and an LD disk wind, namely the treatment
of the lower boundary for the disk. In numerical simulations of 
magnetocentrifugal disk winds the mass flux density from the disk
is given whereas in numerical simulations of LD disk winds 
the mass flux density is a result. The key reason for this difference
is the location of the critical surface at which the mass flux density
is determined. In MHD simulations, this surface corresponds to the 
slow magneto-sonic surface which is located inside the disk, below the
photosphere. Thus unless the disk is included in simulations, 
one  must assume the mass flux density from the lower boundary. 
On the other hand, the critical surface in LD disk winds
is in the supersonic part of the wind above the photosphere.
Therefore, including the photosphere of the disk but not the whole disk
suffices to capture to the transition between the sub-critical and 
super-critical parts
of the LD wind and subsequently  the mass flux density is determined
by the physics of the flow 
(i.e., Feldmeier \& Shlosman 1999).

We have performed test simulations of pure MHD and  MHD-LD disk winds 
following the approach used in studies of magnetocentrifugal winds. 
Our test runs included those where we 
set $v_r=0$, 
$v_\phi=v_{Keplerian}$, $\rho=\rho_0$ and $B_\phi=0$
in the first grid zone above the equatorial 
plane at all times. This setting is very similar to the one
used in numerical simulations of magnetocentrifugal disk winds
(Ouyed \& Pudritz 1997a; 1997b; 1999) with the exception that
they also set the velocity in the direction perpendicular
to the midplane, $v_z$, to some small subsonic velocity 
(i.e., the mass flux density $\rho v_z$ is fixed).
In the test runs, we allow $v_\theta$
to float and let the mass flux density be fixed by the solution
to the problem. 

In short, we found such an approach to modeling MHD and MHD-LD disk winds 
unsatisfactory. For example, in pure MHD cases insisting on a Keplerian
flow in the first grid zone above the equatorial 
plane at all times does not allow proper  modeling
of  magnetic braking of the disk. By
setting $v_r=0$, we prevent or at least significantly reduce
the collapse of the disk  and dragging of magnetic field line 
(e.g., Ouyed \& Pudritz 1997a).
However, when we allowed all variables to float we could successfully
reproduce the evolution of a collapsing disk and an MHD outflow.
In particular, in strong field cases, the disk undergoes dramatic evolution
on a short time scale (i.e., of order of $40~\tau$; in other words, a
couple of orbital periods at $r_\ast$), consistent with the time scale
for magnetic braking of an aligned rotator (Mouschovias \& Paleologou 1980). 

Similarly, we found that the evolution of the disk in a pure MHD
case depends on the conditions in the first grid zone above the equatorial 
plane for weak magnetic fields. We performed several test simulations 
with weak magnetic field, so that the disk was unstable to MRI. In particular,
we adopted $\rho_0=10^{-9} ~\rm g~cm^{-3}$ and $\beta'_w=8\times10^7$ 
yielding $\beta_0=6.4$,
the parameters for which the fastest growing MRI mode is resolved
for our stratified disk. 
For quantities in the first grid zone above the equatorial 
plane set as in magnetocentrifugal models, we found that the disk
evolved very quickly. Initially, we observed a characteristic
exponential growth of the magnetic field, which causes the disk
to separate vertically into horizontal planes. This so-called
channel solution is typical of the development of the axisymmetric
phase of the MRI (Hawley \& Balbus 1991; Goodman \& Xu 1994). 
As expected, the gas streaming is directed  outward in the channel closest 
to the equator and inwards in the channel  higher up from the equator.
However, after about three orbital periods the disk is destroyed.
We find many similarities in the  behavior of the disk 
in our simulations and in previous simulations. 
For example, Stone et al. (1996) and Miller \& Stone (2000)
reported that in their three-dimensional simulations of a 
stratified disk (a local shearing box approximation),
an initially uniform vertical magnetic field leads to a strong radial
streaming in the early phases of the evolution. However, in the late phases, 
high magnetic pressures disrupt the disk and 
the entire computational domain is magnetically dominated, i.e.,
$\beta < 1$ even near the equatorial plane. The high
magnetic pressures are created by strong large-scale radial 
streaming that occurs at large heights because in a stratified disk,
most unstable wavelength increases with height.
As in those local three dimensional simulations, we observed that
strong radial streaming occurs in the horizontal planes into which our disk
separates. Miller \& Stone (2000) concluded that enforcing
Keplerian rotation in the boundary conditions is
appropriate for following the long-term evolution.
We have arrived at the same conclusion, that it is simply impossible to
model on a sufficiently long-time scale an outflow from a disk 
that is quickly disrupted. We found that when
we do not enforce  Keplerian rotation, more precisely when we
allow all the fluid and magnetic field quantities to float, then 
the disk evolves more slowly with radial streaming. 
However when we add the line force and assume too low a density 
the disk can be totally disrupted because of the LD wind.

Motivated by the results from the above test simulations, 
for our standard models, we decided to allow all quantities to float 
in the first grid zone above the equatorial 
plane so that they can evolve self-consistently in all models in Tabel~1.
To ensure we retain a disk for long enough to acquire
a reasonably settled outflow, we
increased  $\rho_0$ from $10^{-9}~\rm g~cm^{-3}$ as in PSD~99
to $10^{-4}~\rm g~cm^{-3}$.
The latter change means that we do not resolve the fastest growing MRI
mode in the disk. To do so we would have to increase
the strength of the magnetic field for a given 
resolution  or increase the resolution. Unfortunately, we cannot afford 
either of these modifications
because they would make our already computationally
demanding simulation even more demanding.
Therefore, in this first attempt to model MHD-LD disk winds
we resolve  the wind from the disk but do not resolve 
the disk itself. Thus we cannot resolve the fastest growing 
MRI mode within the disk scale height, but only 
most unstable modes at larger  heights.

We finish with a remark that in the literature 
most of the numerical simulations
of MHD disk winds were stopped after one or two or at most a few
orbital periods. The reason for this is obvious but it is 
disappointing that one cannot perform yet global simulations
of magnetized disks, treating properly accretion as well as outflow
and choosing physical parameters comparable to those in real systems.

\subsection{Limitations of models and future work}

The most important limitation of our model is an inadequate spatial
resolution for modeling the MRI inside the disk 
(see discussion above).  As is fitting for a first exploration of MHD-LD
wind  models from  disks, we aim to examine only the parameter space of 
our models that will define the major trends in disk wind behavior. 
Therefore our priority is to set up the simulation in such a way that 
the base of the wind is relatively stable and corresponds to a steady state 
accretion disk. Obviously the problem with modeling MRI disappears when 
the magnetic field is strong. In the context of MHD winds, a strong
magnetic field case for which the disk is MRI-stable 
is also  physically interesting.

We have explored  MHD-LD models where  initially $\beta < 1$
everywhere on the grid. However, we found then Alfv${\acute{\rm e}}$n speeds
so high that the resulting time step was extremely small.
We emphasize that our choice of $\beta'_w$ is limited by the constraints
on the density in the computational domain, i.e.,
the need to have a relatively large contrast between the density
near the equatorial plane and that high above  the plane. 
Increasing arbitrarily the lower limit, $\rho_{min}$, to reduce 
the density contrast and subsequently reduce the Alfv${\acute{\rm e}}$n speed
is not suitable  to modeling LD winds. The line force is very sensitive to 
the density and our wind solution can then include gas with a spurious 
density set by the lower limit.
Thus we were very cautious in choosing $\rho_{min}$, so that
the region with $\rho \le \rho_{min}$ 
has no effect on the disk flow.

We note that other numerical simulations explored
a relatively low density contrast between the disk and the ambient
gas (e.g., in Stone \& Norman 1994; Ouyed \& Pudritz 1997a, 1997b, 1999;
and Krasnopolsky, Li \& Blandford 1999 the density contrast is $\simless
10^3$). 
Here we deal with density contrasts several orders of magnitude higher. 

The fact that our results strongly depend on the magnetic field
points to a need to explore different configurations for the initial
magnetic field and 
to move from two-dimensional axisymmetric simulations
to fully three-dimensional simulations.
We are interested in the long-time evolution of the flow. Therefore
three-dimensional simulations are required as there exist no self-sustained
axisymmetric dynamos. Thus, contrary to the outflows from stars,
simulations of outflows from magnetized disks -- with or without radiation
pressure -- should include the disks themselves, not just
the disk photosphere, and should be performed
in three dimensions.

\section{Conclusions}

We have studied winds from accretion disks with magnetic fields and 
the radiation force due to lines. We use numerical methods to solve 
the two-dimensional, time-dependent equations of  ideal MHD. We have 
accounted for the radiation force using a generalized multidimensional 
formulation of the Sobolev approximation. For the initial conditions,
we have considered uniform vertical magnetic fields and geometrically
thin, optically thick Keplerian disks. We allow the magnetic field
and the disk to evolve. In particular, we do not enforce  Keplerian 
rotation in the first grid zone above the equatorial plane. Although the gas 
near the  equatorial plane departs only slightly from  Keplerian rotation
in our self-consistent calculations, 
its evolution is notably different when we enforce  Keplerian rotation.
We find  that the magnetic fields very quickly start deviating from purely 
vertical due to the MRI. This leads to fast growth of the toroidal magnetic 
field as field lines wind up due to the disk rotation. As a result, 
the toroidal field dominates over the poloidal field above  the disk and 
the gradient of the former drives a slow and dense disk outflow,
which conserves specific angular momentum.
Depending on the strength of the magnetic field relative to the system 
luminosity, the disk wind can be radiation- or MHD driven. 
For example, for our model parameters $\MDOT_a~=~10^{-8}~\MSUNYR$ and $x=0$, 
the wind is radiation-driven for $\beta'_w\simgreat 1$ and, as   
its pure LD counterpart, consists of a dense, slow outflow that 
is bounded on the polar side by a high-velocity  stream. 
The mass-loss rate is mostly due to the fast stream. 
As the magnetic field strength increases (i.e., $\beta'_w <10^{-1}$),  
first the slow part of the flow is affected. In particular,
the slow wind  becomes  denser  and faster than its pure LD 
counterpart by a factor of $\simgreat 100$ and $\simgreat 3$, respectively.
Consequently, the dense wind begins to dominate  the mass-loss rate. 
In very strong magnetic field (i.e., $\beta'_w \simless 10^{-3}$) 
or pure MHD cases, the wind consists of only 
a dense, slow outflow without the presence of the distinctive fast 
stream so typical of pure LD winds. Our simulations indicate 
that winds launched by magnetic fields are likely to remain dominated 
by the fields downstream because of their relatively high densities. 
The radiation force due to lines may not be able to change a dense MHD wind 
because the line force strongly decreases with the density.

Our results show that, as expected, a hybrid model predicts mass-loss rates 
higher than those predicted by a pure LD model. 
We plan to compute synthetic  line profile based on our simulations
and check whether our MHD-LD models can resolve the problem of nMCV winds.
However even now we can say that our MHD-LD models may not solve the problem
because 
to explain nMCV winds we need a model that predicts not only 
a higher $\MDOT_w$ than a LD wind model but also $\MDOT_w$ must be mostly 
due to a fast wind not a dense slow wind as we find our models
(Proga et al. 2002).

ACKNOWLEDGMENTS: We thank J.M. Stone, M.C. Begelman, J.E. Drew, P.J. Armitage, 
A. Feldmeier, and M. Ruszkowski for useful discussions.
In particular, we thank A. Feldmeier for sharing with us his results
and conclusions on MHD-LD disk winds. 
We also thank  an anonymous referee for comments
that helped us clarify our presentation.
We acknowledge support from NASA under LTSA grant NAG5-11736.
We also acknowledge support provided by NASA through grant  AR-09532
from the Space Telescope Science Institute, which is operated 
by the Association of Universities for Research in Astronomy, Inc., 
under NASA contract NAS5-26555.
Computations were partially supported by NSF grant AST-9876887.

\newpage
\section*{ REFERENCES}
 \everypar=
   {\hangafter=1 \hangindent=.5in}

{

  Abbott, D.C. 1982, ApJ, 259, 282

  Balbus, S.A., \& Hawley, J.F. 1991, ApJ, 376, 214

  Balbus, S.A., \& Hawley, J.F. 1998, Rev. Mod. Phys., 70, 1 

  Batchelor G.K. 1967, An Introduction to Fluid Mechanics (Cambridge:
  Cambridge University Press)

  Blandford R.D., Payne D.G. 1982, MNRAS, 199, 883

  Cannizzo J. K., Pudritz R.E. 1988, ApJ, 327, 840

  Castor J.I., Abbott D.C.,  Klein R.I. 1975, ApJ, 195, 157 (CAK)
  
  Contopoulos J. 1995, ApJ, 450, 616

  Chandrasekhar, S. 1960, Proc. Nat. Acad. Sci., 46, 253

  Drew J.E., Proga D., in  {\it  ``Cataclysmic Variables'',
  Symposium in Honour of Brian Warner}, Oxford 1999, 
  ed. by P. Charles,  A. King, D. O'Donoghue, in press   

  Feldmeier, A., \& Shlosman, I. 1999, ApJ, 526, 344

  Feldmeier, A.,  Shlosman, I., \&, Vitello, P. 1999, ApJ, 526, 357

  Gayley, K.G. 1995, ApJ, 454, 410

  Goodman J.,  Xu G. 1994 ApJ, 432, 213

  Hawley J.F,  Balbus S.A. 1991, ApJ, 376, 214

  Kato, S.X., Kudoh T., \& Shibata K. 2002, ApJ, 565, 1035

  K$\ddot{\rm o}$nigl A. 1993, in  {\it ``Astrophysical Jets''},
  ed. by D.P. O'Dea (Cambridge: Cambridge Univ. Press), 239
  
  Krasnopolsky R., Li Z.-Y., Blandford R., 1999, ApJ, 526, 631 

  Kudoh, T., Matsumoto, R., \& Shibata, K. 1998, ApJ, 508, 186

  Kudoh T., Shibata K. 1997, ApJ, 474, 362

  Li, Z.-Y. 1995, ApJ, 444, 848

  Li, Z.-Y. 1996, ApJ, 465, 855

  Miller, K.A., Stone, J.M. 2000, ApJ, 534, 398

  Mouschovias, T. Ch., Paleologou, E.V. 1980, ApJ, 237, 877

  Ogilvie G.I., Livio M. 1998, ApJ, 499, 329

  Ouyed R., Pudritz R.E. 1997a, ApJ, 482, 717

  Ouyed R., Pudritz R.E. 1997b, ApJ, 484, 794

  Ouyed R., Pudritz R.E. 1999, MNRAS, 309, 233

  Owocki S.P., Cranmer S.R.,  Gayley K.G.  1996, ApJ, 472, L115 

  Pelletier G., Pudritz R.E. 1992, ApJ, 394, 117

  Pereyra N.A., Kallman T.R. Blondin J.M. 1997, ApJ, 477, 368

  Pereyra N.A., Kallman T.R. Blondin J.M. 2000, ApJ, 532, 563

  Proga D. 1999, MNRAS, 304, 938

  Proga D. 2000, ApJ, 538, 684

  Proga D. 2002, Mass Outflow in Active Galactic Nuclei: New Perspectives, 
  ASP Conference Proceedings, Vol. 255. 
  Edited by D. M. Crenshaw, S. B.
  Kraemer, and I. M. George. ISBN: 1-58381-095-1. 
  San Francisco: Astronomical Society of the Pacific, 2002., p.309

  Proga, D., \& Kallman, T.R. 2002, ApJ, 565, 455

  Proga, D., Kallman, T.R., Drew, J.E., \& Hartley, L.E., 2002, ApJ, 572, 382

  Proga D., Stone J.M., Drew J.E. 1998, MNRAS, 295, 595 (PSD~98)

  Proga D., Stone J.M., Drew J.E. 1999, MNRAS, 310, 476  (PSD~99)

  Proga D., Stone J.M., Kallman T.R. 2000, ApJ, 543, 686  

  Pudritz R.E., Norman C.A. 1986, ApJ, 301, 571 

  Romanova, M.M., Ustyugova, G.V., Koldoba, A.V., Chechetkin, V.M., \&
  Lovelace, R.V.E. 1997, ApJ, 482, 708

  Shakura N.I., Sunyaev R.A. 1973 A\&A, 24, 337

  Shibata K.,  Uchida Y. 1986, PASJ, 38, 631

  Stone, J.M., Hawley, J.F., Gammie, C.F., \& Balbus, Steven A. 1996 ApJ,
  463, 656

  Stone J.M., Norman M.L. 1992a, ApJS, 80, 753

  Stone J.M., Norman M.L. 1992b, ApJS, 80, 791
  
  Stone J.M., Norman M.L. 1994, ApJ, 433, 746

  Uchida Y., Shibata K. 1985, PASJ, 37, 515

  Ustyugova G.V., Koldoba A.V., Romanova M.M. Chechetkin V.M. Lovelace R.V.E.
  1995, ApJ, 439, 39L

  Ustyugova G.V., Koldoba A.V., Romanova M.M. Chechetkin V.M. Lovelace R.V.E.
  1999, ApJ, 516, 221
  
  Velikov, E.P. 1959, JETP, 36, 1398

  Vitello P.A.J.,  Shlosman I., 1988 ApJ, 327, 680
}

\newpage
Fig.~1. A sequence of density maps (left), velocity fields (middle)
and contours of the toroidal magnetic field (right)
from run C0D after 37.8, 166.2, 294.7, 423.1, and 551.6~$\tau$ 
(time increases from top to bottom).  
Run~C0D is the example of a MHD-LD flow discussed in detail in section 3.1.
For clarity, we suppress velocity vectors for regions with very low
density (i.e., $\rho < 10^{-15} \rm g~cm^{-3}$).  
The solid line overplotted on the velocity maps marks the 
the Alfv${\acute{\rm e}}$nic surface 
(i.e., location where $\mid v_{Ap} \mid= \mid v_p \mid$, middle panels).
The $B_\phi$ contours are for -30,-20,-10,-5.,-1, 0, and 5.
Dotted lines denote negative values of $B_\phi$.

Fig.~2. Two-dimensional structure of several quantities from
model C0D after 680~$\tau$.
The contours of $L$ are for 1.0, 1.5, 2 ,2.5 , and 3.0 
while for $\Omega$ are for 0.025, 0.05, 0.1, 0.2, 0.4, and 0.8.
Both $L$ and $\Omega$ are in units of the specific angular momentum
and angular velocity on the Keplerian disk at $r_\ast$.
The contours of $log \beta$ are for -3, -2.5 , -2.0, and -1.0.
The solid lines overplotted on the poloidal magnetic field (bottom
middle panel)
show the location where the strength of the poloidal magnetic 
field equals  the toroidal magnetic field, $\mid B_p \mid =\mid B_\phi
\mid$. The contours for $B_\phi$ are as in Figure~1.

Fig.~3. Quantities at the outer boundary in model C0D after 680~$\tau$.
The ordinate on the left hand side of each panel refers to the solid line, 
while the ordinate on the right hand side refers to the dotted line.
Of particular note is the continuous strong increase of  
the accumulated  mass loss rate with increasing polar angle.
This is associated with the fact that the mass loss rate is dominated
by a slow dense wind originating at large radii.

Fig.~4. Model mass loss rates as functions of the total Eddington factor
(top panel) and the initial plasma parameter, $\beta'_w$
(bottom panel). In the top panel,  open circles are for pure LD models, 
and the other shapes correspond to different values of $\beta'_w$ 
($\beta'_w=2\times10^{0}$ crosses, 
 $\beta'_w=2\times10^{-2}$ asterisks, 
 $\beta'_w=2\times10^{-3}$ diamonds, 
 $\beta'_w=2\times10^{-4}$ triangles,  and
 $\beta'_w=2\times10^{-6}$ squares).
Table~1 specifies other model parameters.  
The thick solid line represents results in the stellar CAK
case while the thin solid line represents  the stellar CAK
case corrected for a finite value of the maximum force multiplier
$M_{max}$ (see the text for more detail).
The alternative ordinate on the
right hand side of the lower panel is the dimensionless wind mass loss
rate parameter $\dot{M}_{w}'$ defined in PSD~98 (see equation 22 in PSD~98).
The solid line in bottom panel represents  the mass loss rate
for models of series C0  while the dashed line represents  the mass loss rate
for models of series D1 (see section 3.2).

\newpage

\begin{figure}
\begin{picture}(280,590)
\put(100,480){\includegraphics{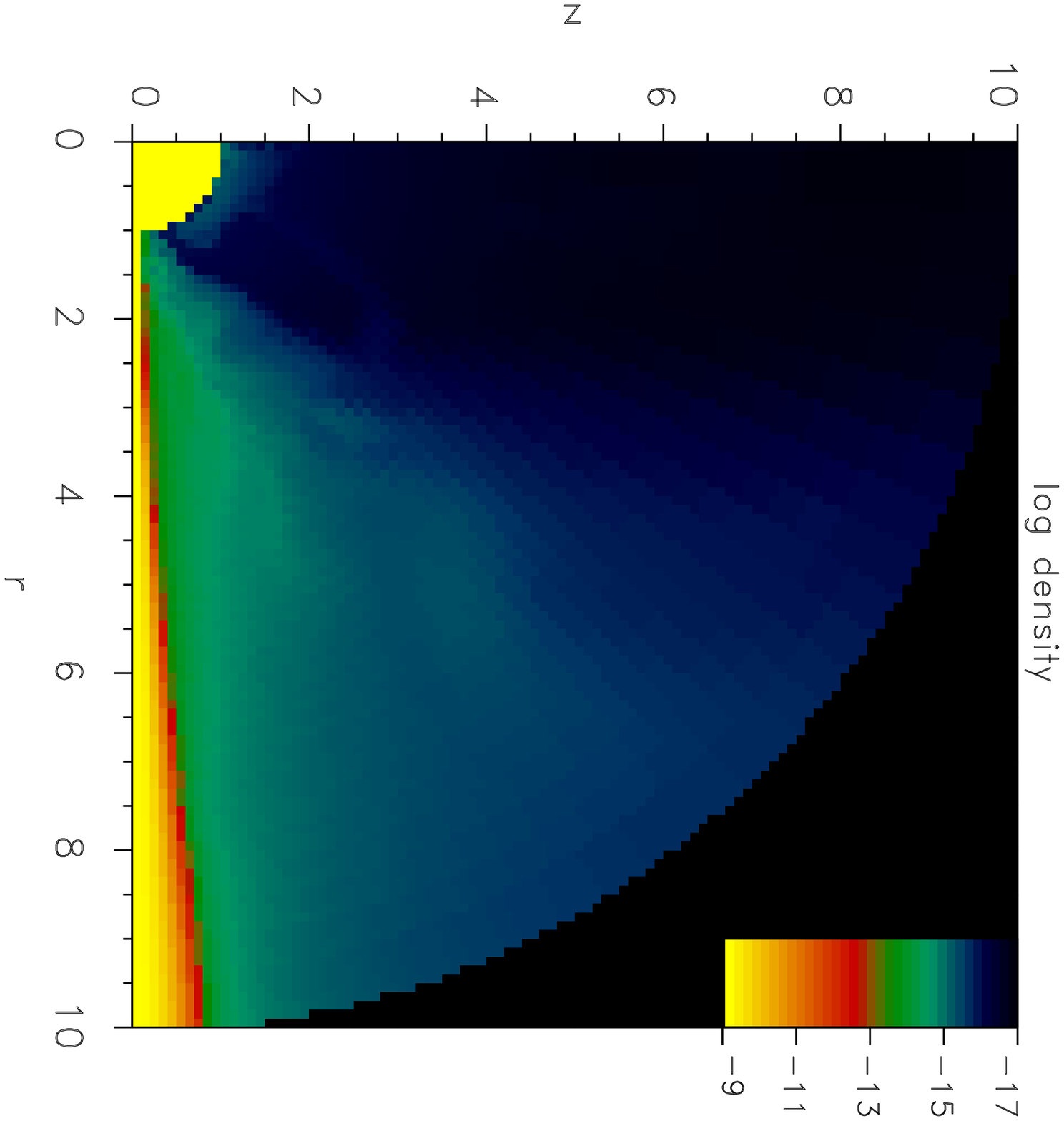}}
\put(100,360){\includegraphics{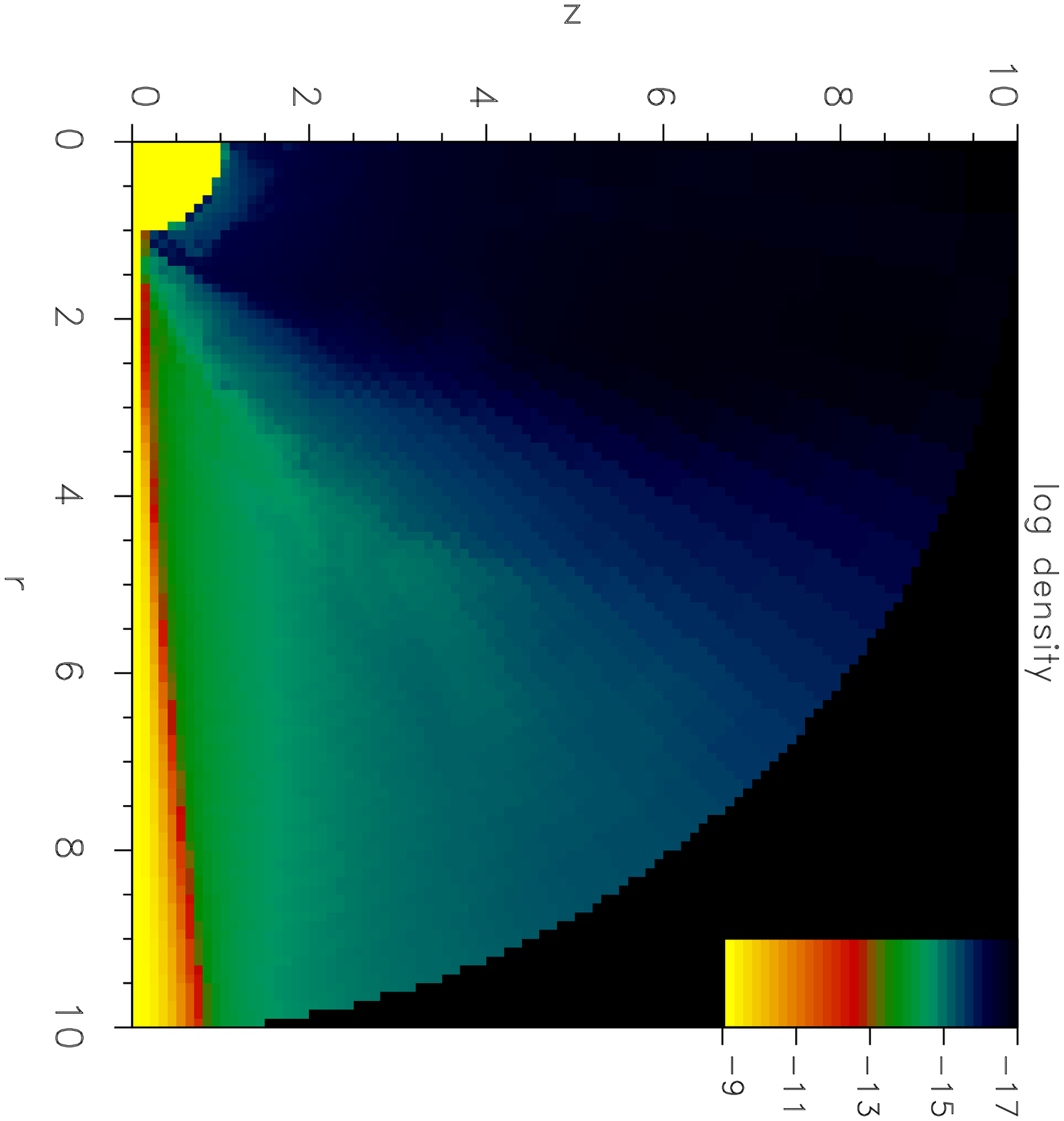}}
\put(100,240){\includegraphics{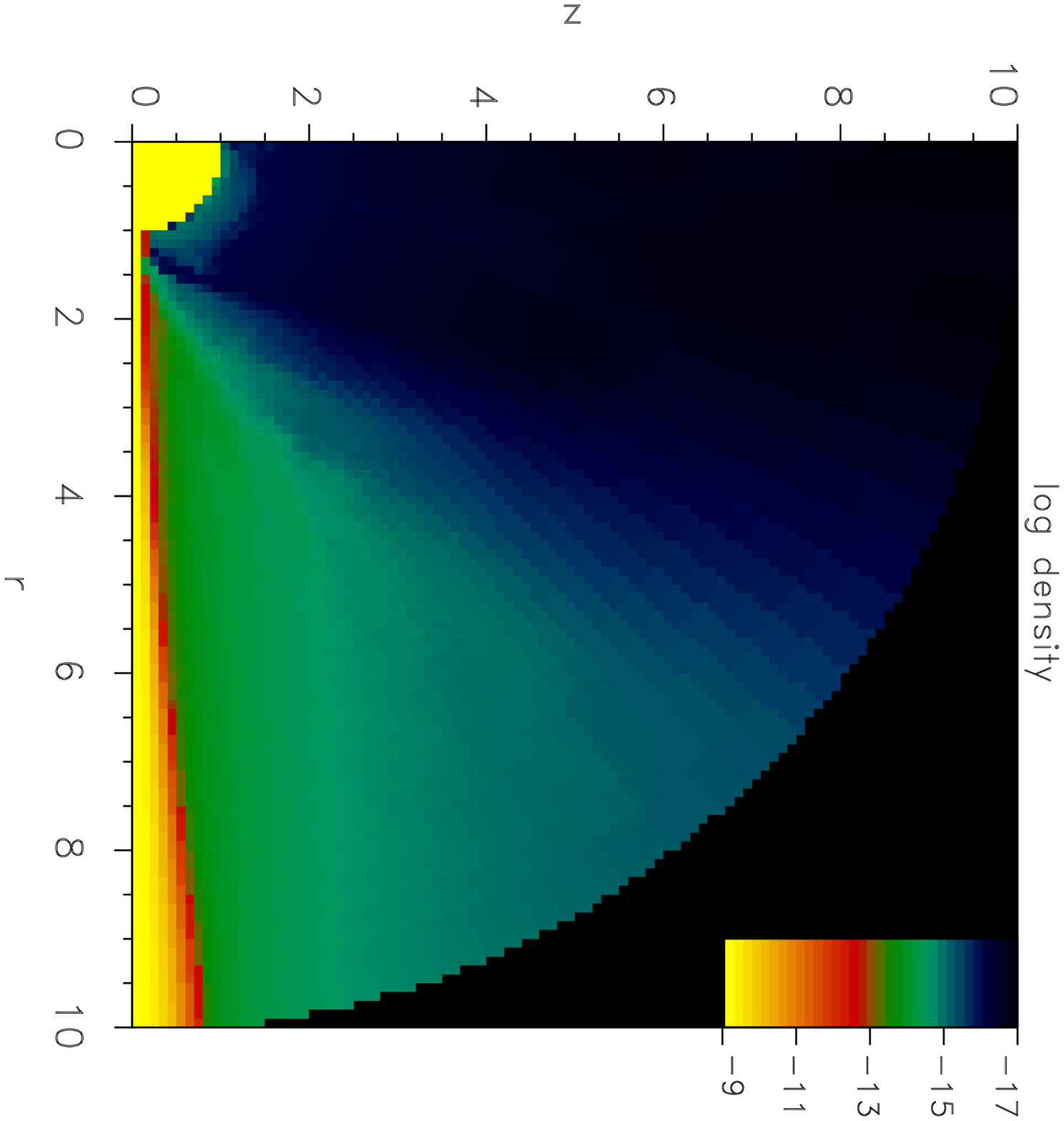}}
\put(100,120){\includegraphics{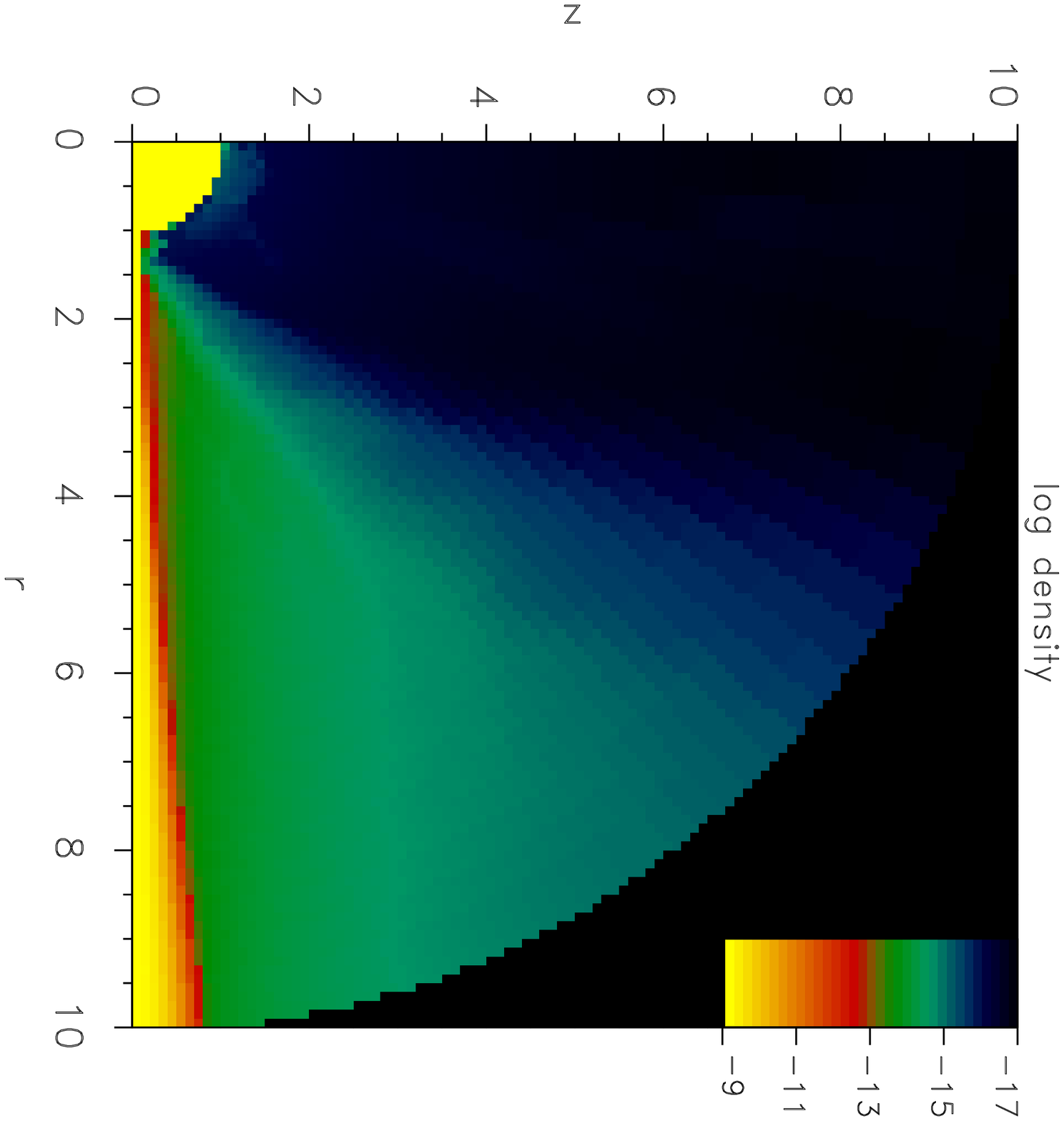}}
\put(100,0){\includegraphics{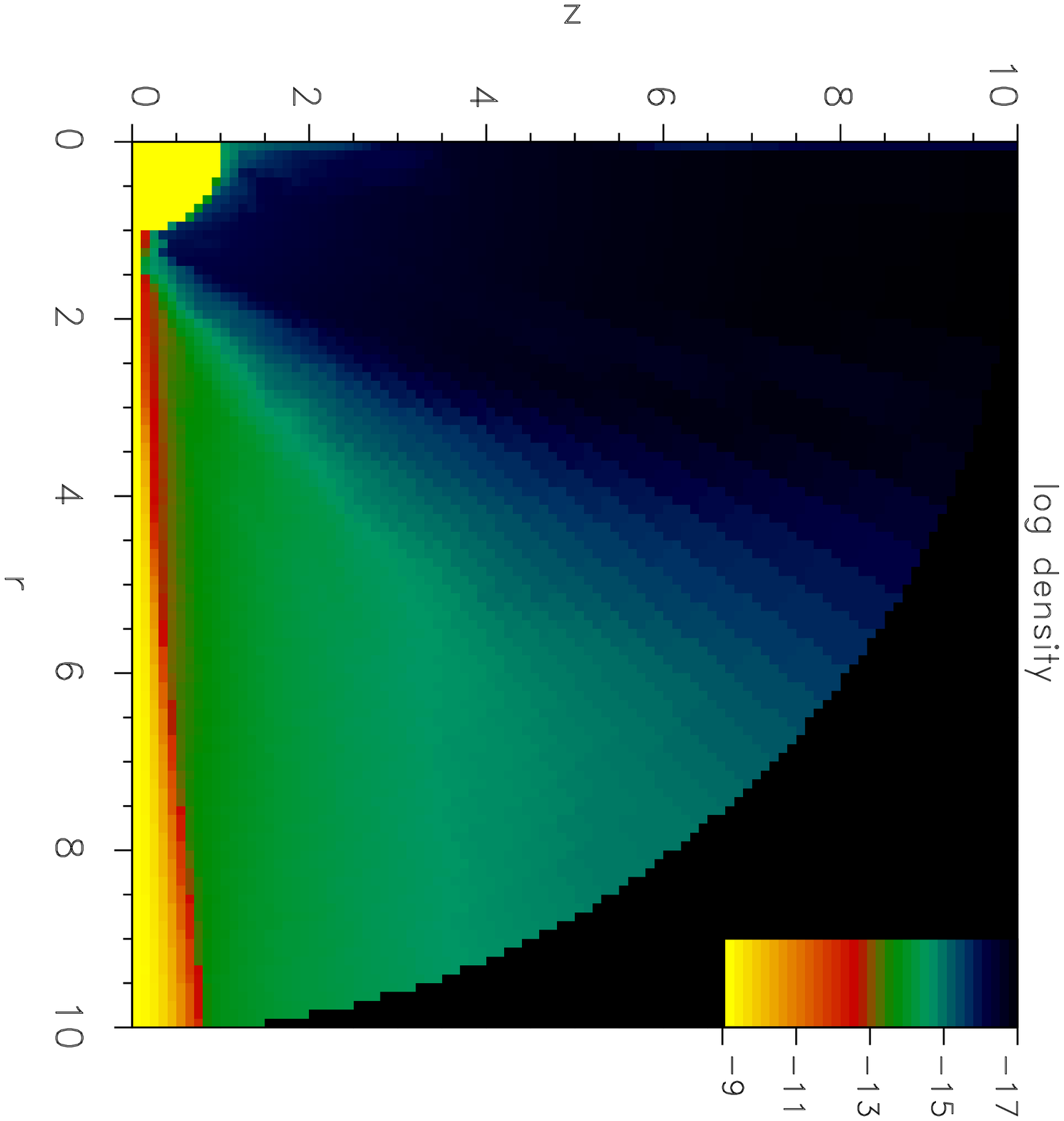}}

\put(250,480){\includegraphics{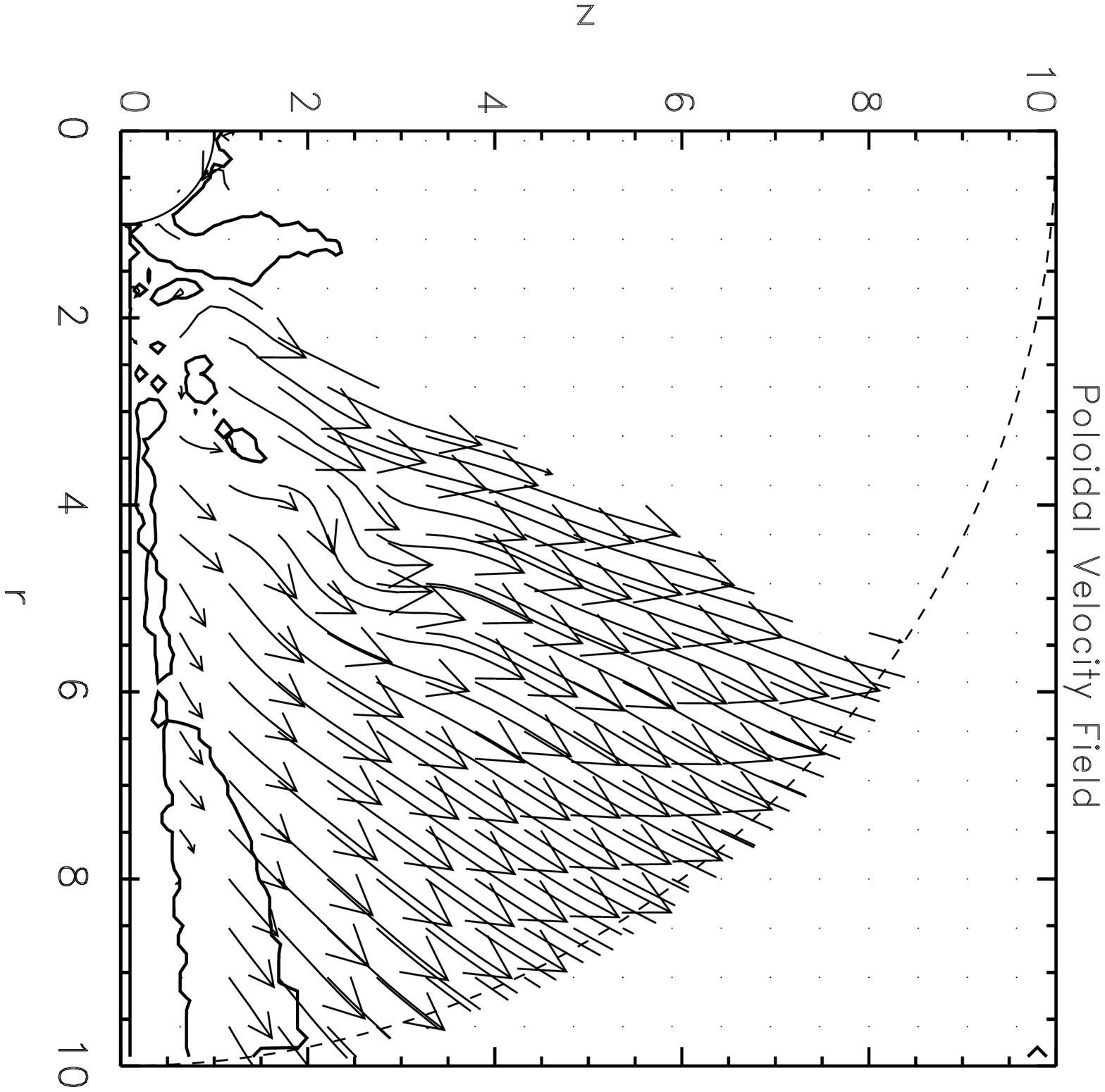}}
\put(250,360){\includegraphics{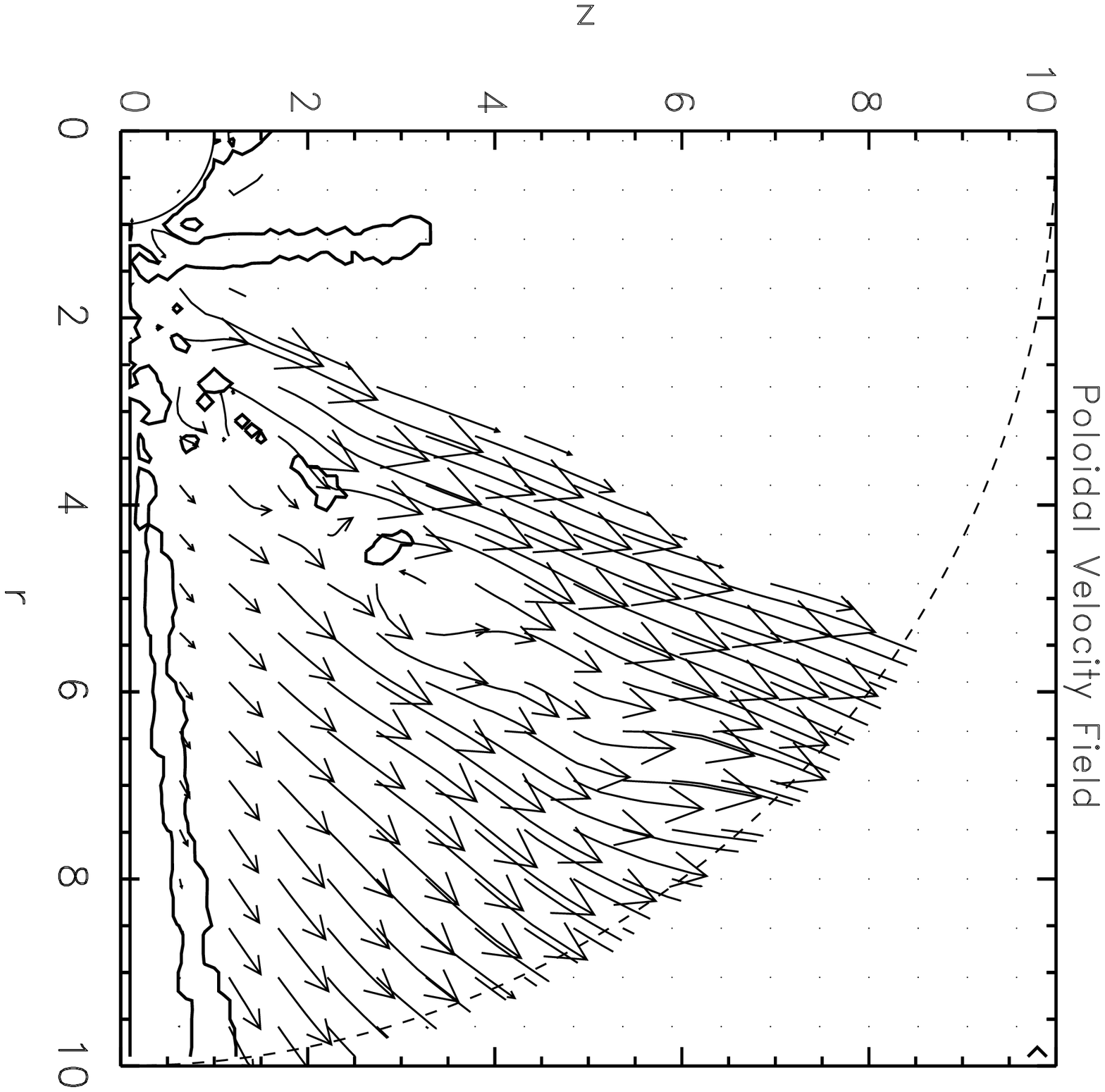}}
\put(250,240){\includegraphics{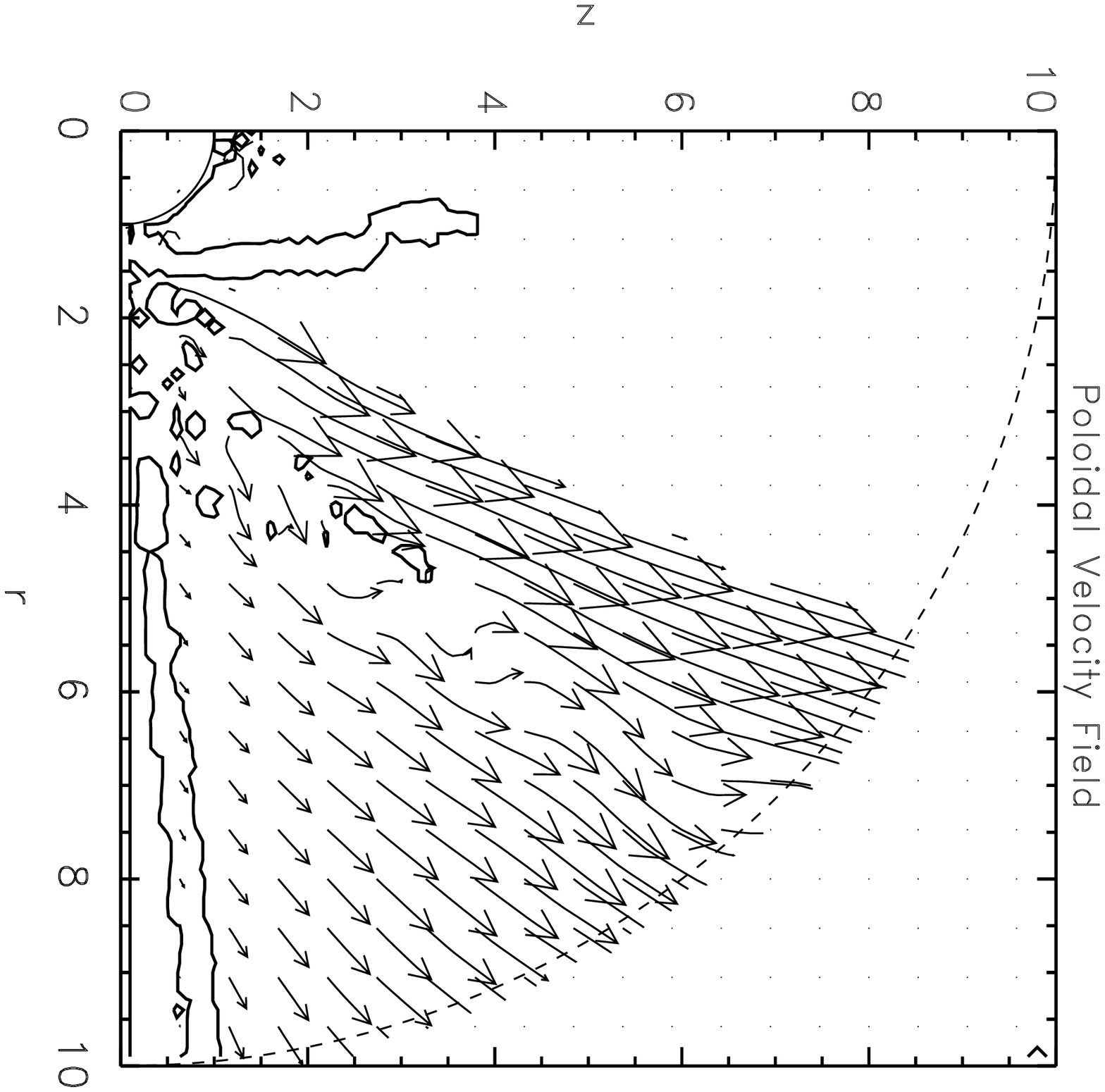}}
\put(250,120){\includegraphics{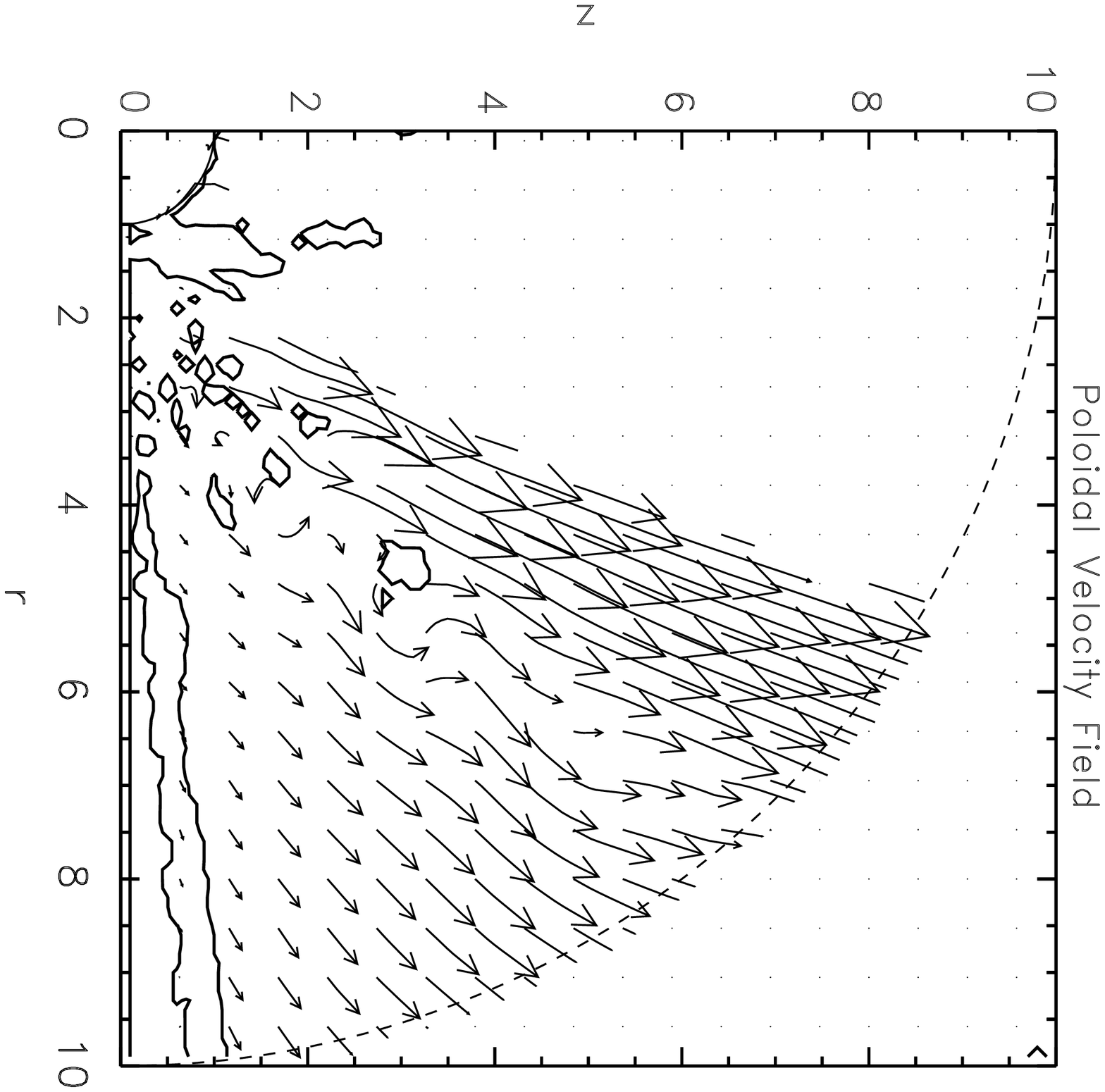}}
\put(250,0){\includegraphics{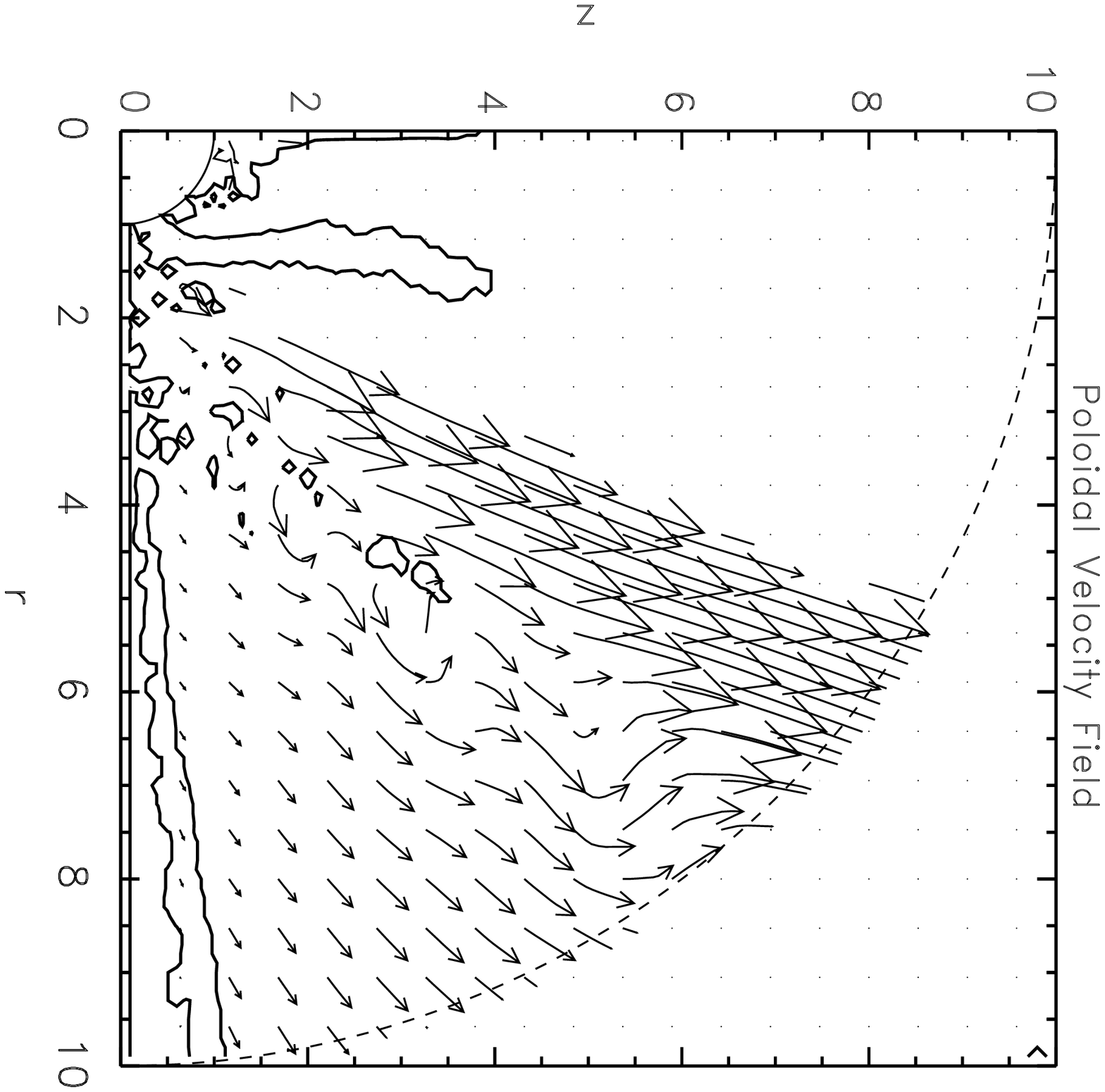}}

\put(400,480){\includegraphics{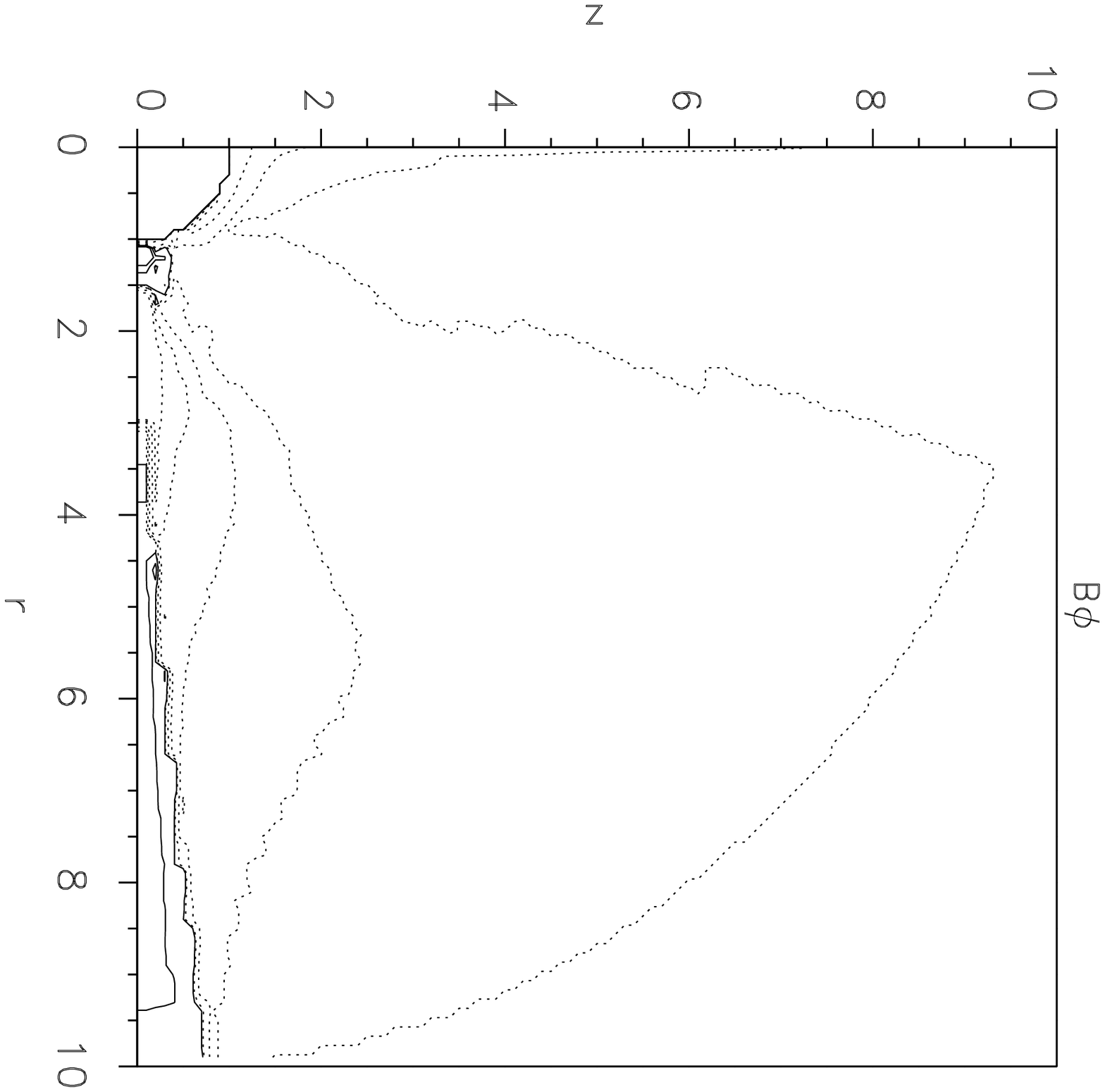}}
\put(400,360){\includegraphics{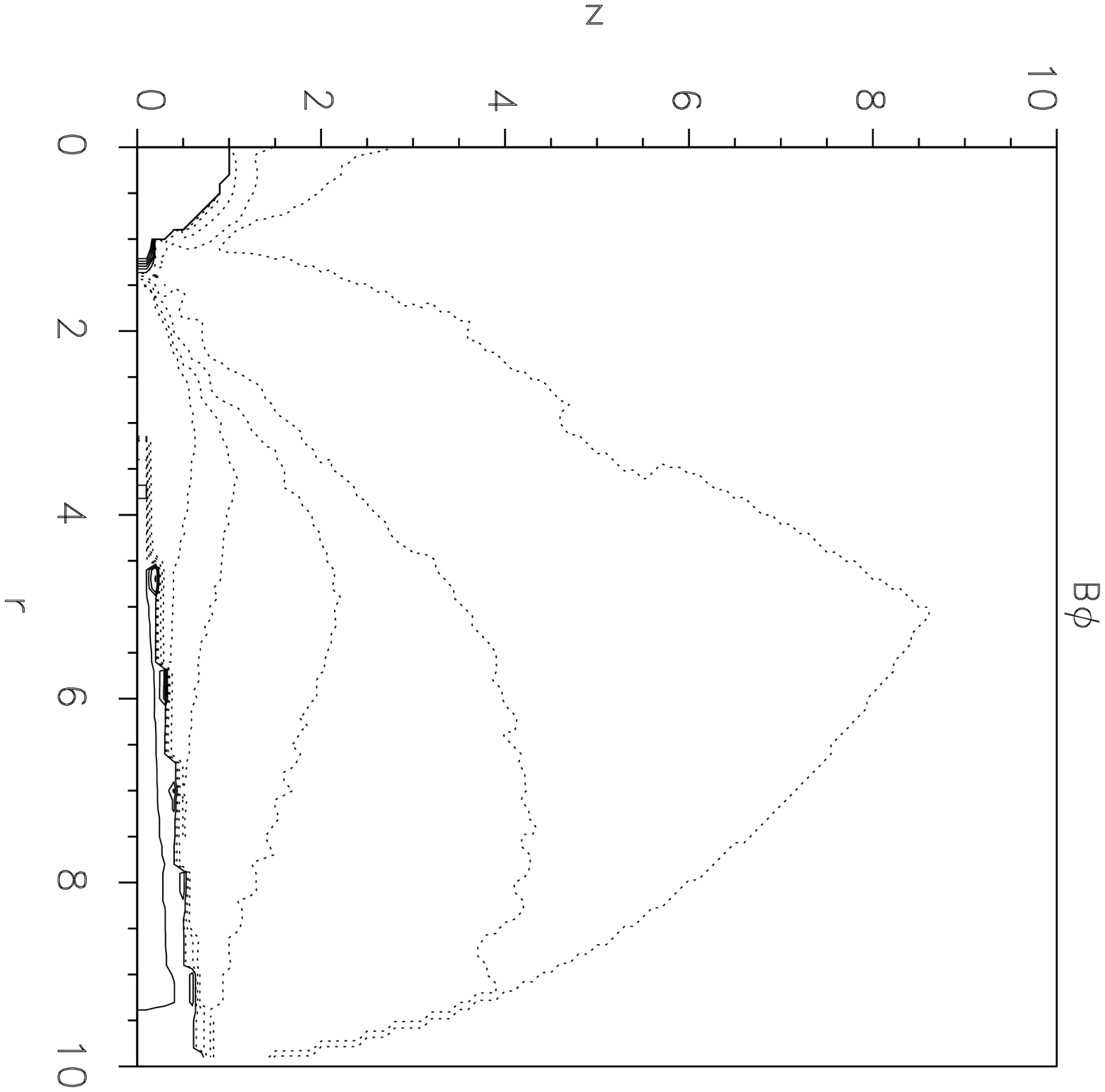}}
\put(400,240){\includegraphics{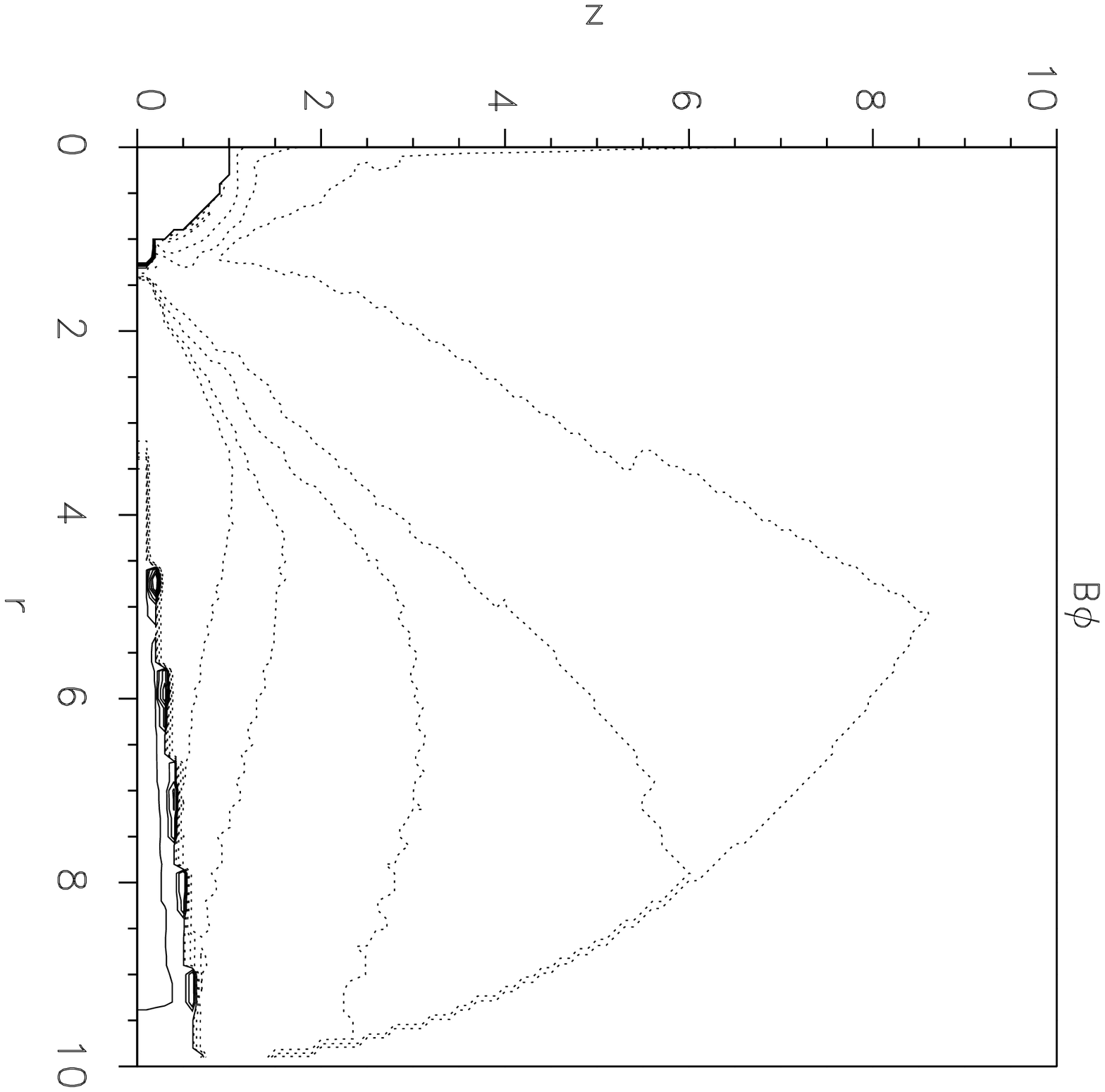}}
\put(400,120){\includegraphics{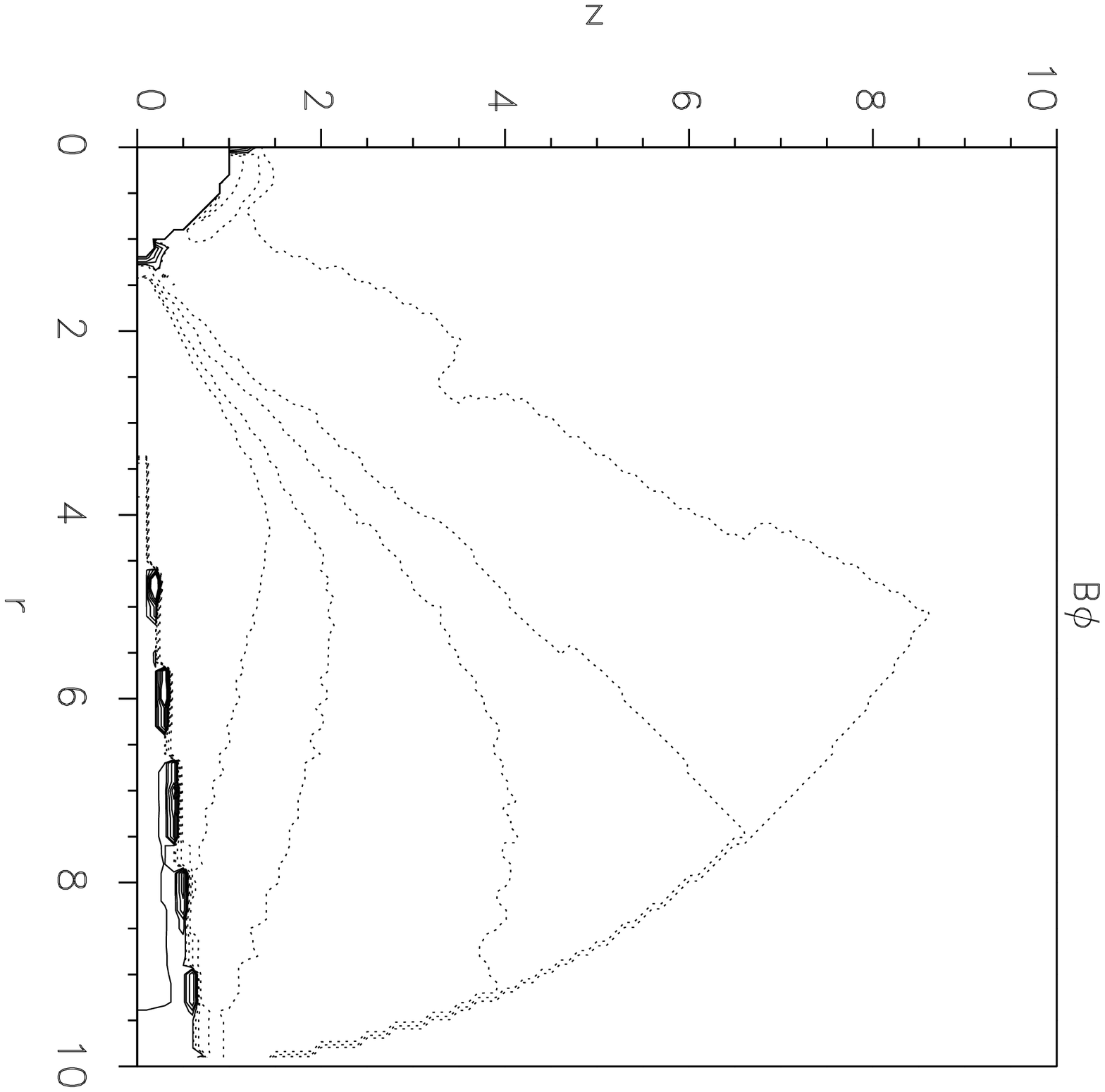}}
\put(400,0){\includegraphics{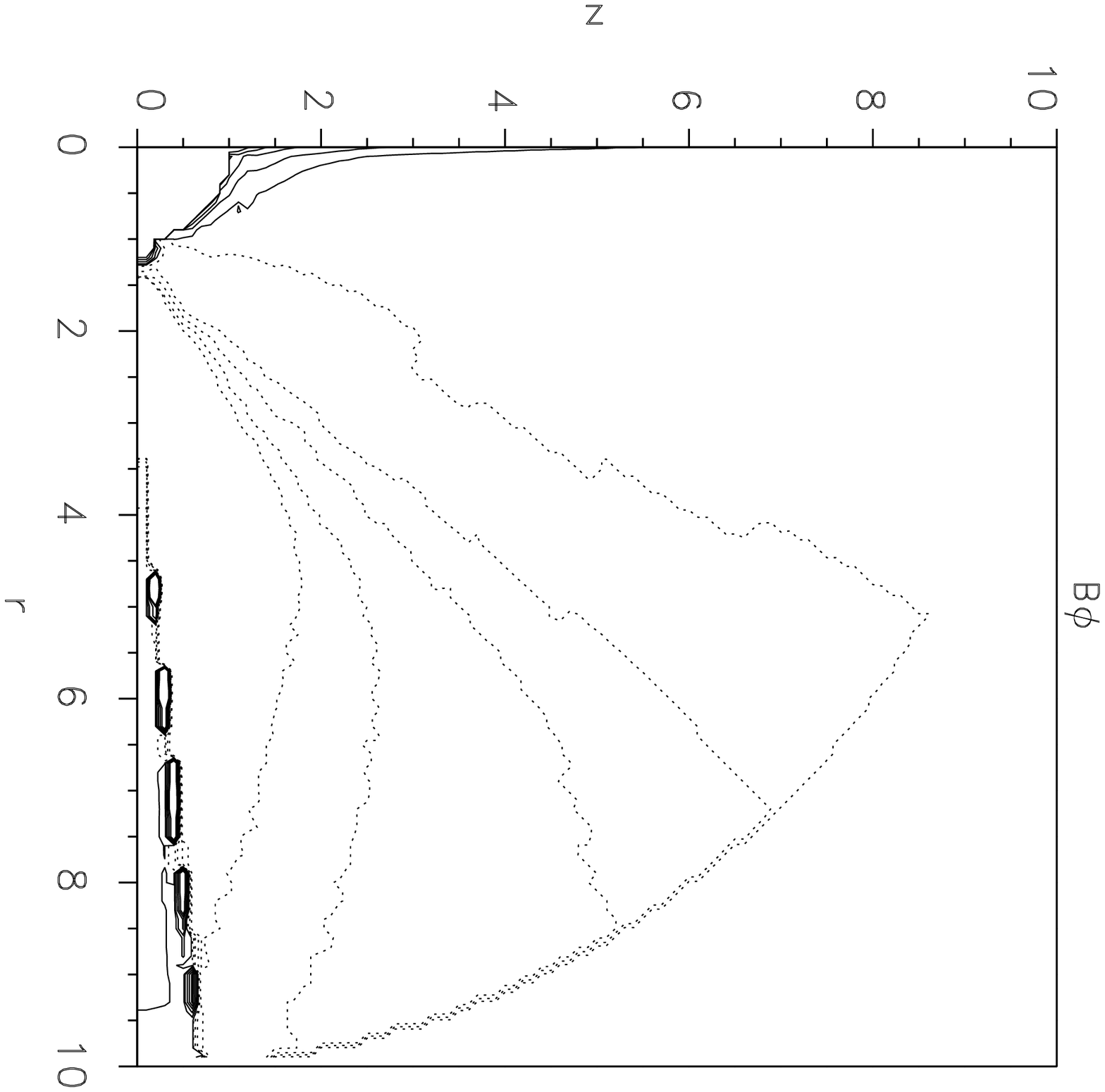}}

\end{picture}
\caption{ 
}
\end{figure}

\begin{figure}
\begin{picture}(180,400)
\put(50,0){\includegraphics{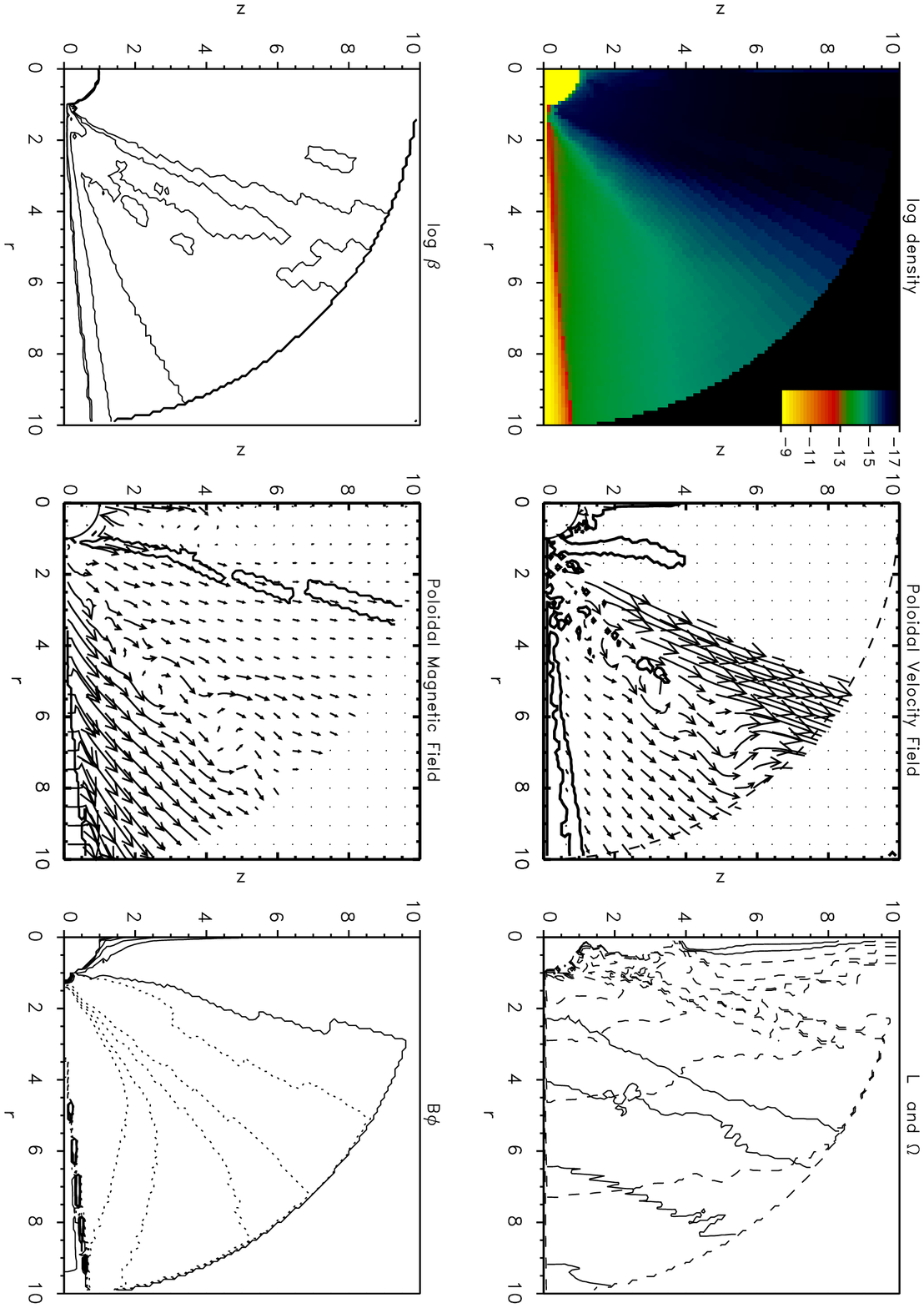}}
\end{picture}
\caption{ 
}
\end{figure}

\begin{figure}
\begin{picture}(180,400)
\put(50,0){\includegraphics{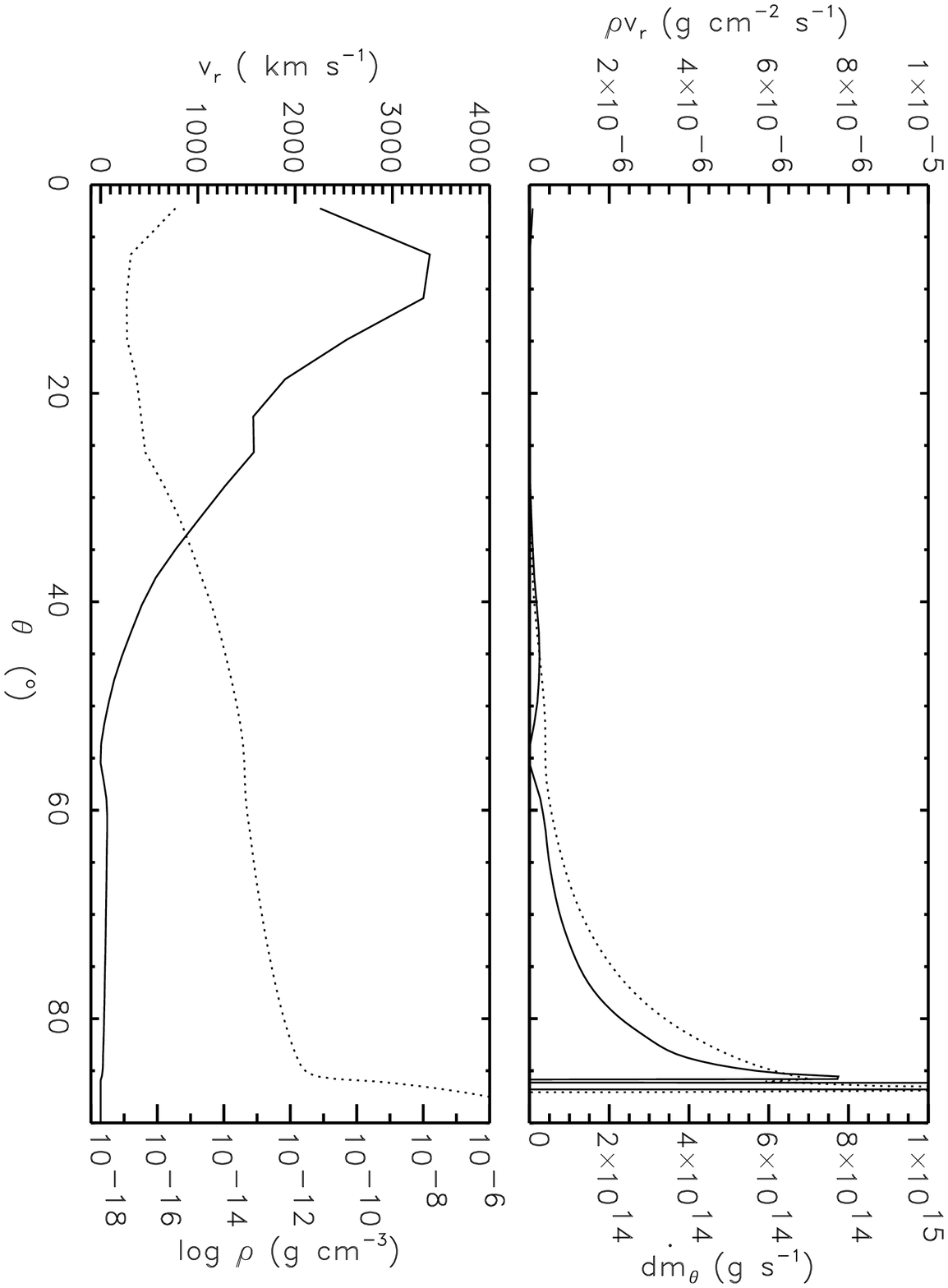}}
\end{picture}
\caption{ 
}
\end{figure}

\begin{figure}
\begin{picture}(280,590)
\put(50,200){\includegraphics{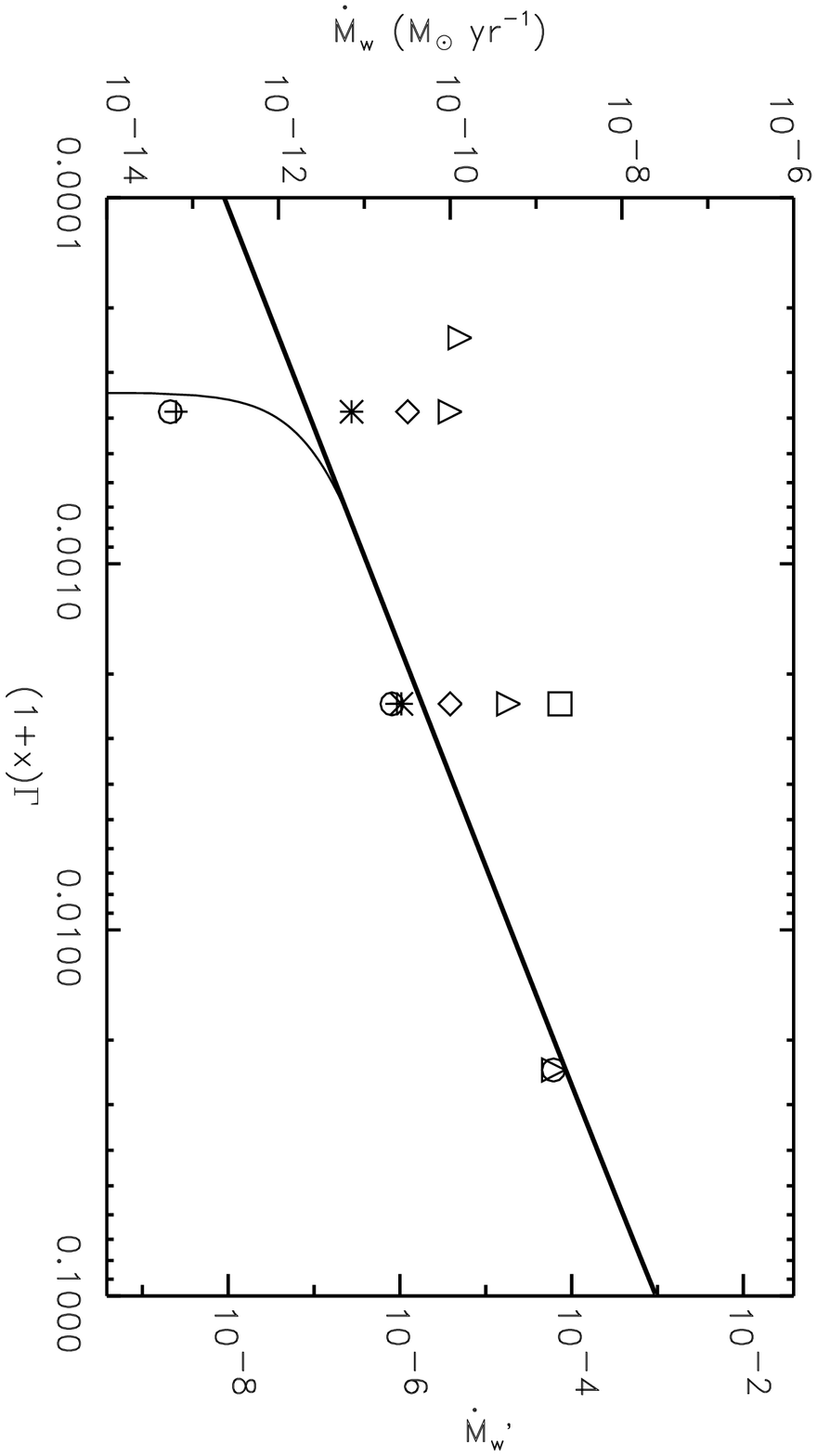}}

\put(50,-100){\includegraphics{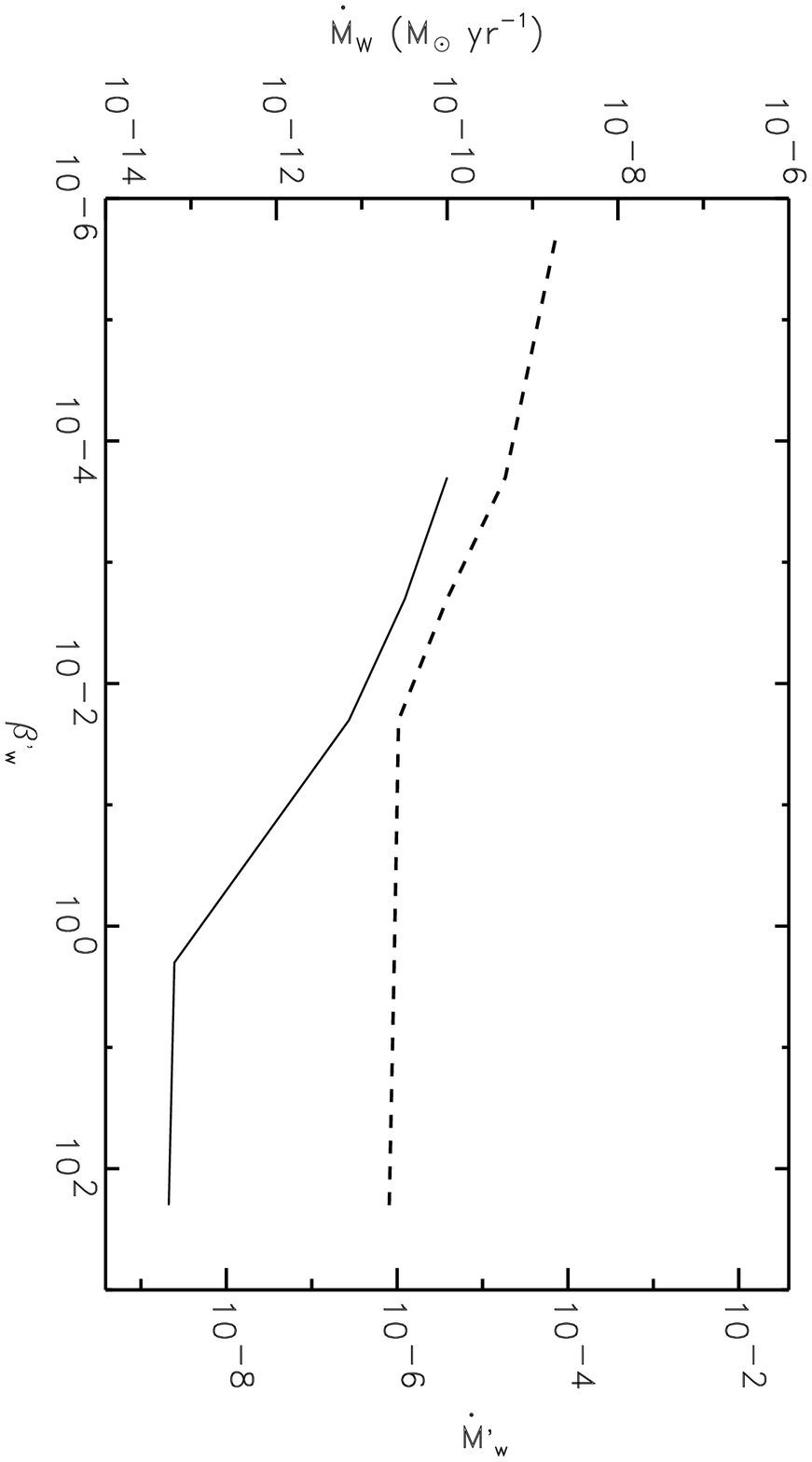}}
\end{picture}
\caption{ 
}
\end{figure}

\eject

\normalsize
\centerline{ Table 1. Summary of results for disc winds }
\begin{center}
\begin{tabular}{c c c c c c c c l } \\ \hline \hline
     &                         &  & &                         &                   &             \\
Run$^\ast$ &  $\MDOT_a$ & $x$ & $\beta'_{w}$& $t_f$ & $\MDOT_w$ & $v_r(10 r_\ast)$ & $\omega$ & comments \\
     & (M$_{\odot}$ yr$^{-1}$ ) &   & & $\tau$ & (M$_{\odot}$ yr$^{-1}$) &   
$(\rm km~s^{-1})$ & degrees  &  \\  

\hline

pure LD     &                         &   & &      &                   &                   &            & \\

C0A      &$  10^{-8}$                & 0 & $\infty$ & 800 &$ 5.5\times10^{-14}$     & 900                 &  50 & run~A in PSD~99\\
 & & & & &  & \\
B1A    &$ \pi \times 10^{-9}$     & 1 & $\infty$ & 1300 &                         &                 &   & no supersonic outflow \\
D1A    &$ \pi \times 10^{-8}$    & 1 & $\infty$ & 1000&$ 2.1\times10^{-11}$    & 3500                &  32 &  run~C in PSD~99\\
E1A     &$ \pi \times 10^{-7}$    & 1 & $\infty$ & 1000&$ 1.6\times10^{-9}$     & 7000                &  32 &  \\

 & & & & &  & & \\

pure MHD     &                         &   &      &                   &                   &           &  &\\
A0C     &$  0 $                    & 0 & $2\times10^{-1}$ &1500 &$6.4\times10^{-13}$     & 100                 &  65 & \\
 & & & & &  &  \\
LD-MHD     &                         &   &      &       &            &                   &           &  \\
C0B    &$  10^{-8}$              & 0 & $2\times10^0$ & 400 &$ 6.4\times 10^{-14}$     & 900                 &  55 & \\
C0D    &$  10^{-8}$              & 0 & $2\times10^{-2}$ & 680&$ \simgreat 7.1\times10^{-12}$     & 1000                 &  60 & \\
C0E    &$  10^{-8}$              & 0 & $2\times10^{-3}$ & 450&$>3.2\times10^{-11}$     & 1000                 &  65 & \\
C0F    &$  10^{-8}$              & 0 & $2\times10^{-4}$ & 170&$ \simgreat 1.0\times10^{-10}$     & 1300                 &  65 & \\
 & & & & &  & & \\
B1F    &$ \pi \times 10^{-9}$    & 1 & $2\times10^{-4}$ & 300 &$1.3\times10^{-10}$     & 600                 & 70  &    \\
 & & & & &  & \\
D1B    &$ \pi \times 10^{-8}$    & 1 & $2\times10^0$ &250 &$ 2.4\times10^{-11}$     & 3500                &  32 &  \\
D1D    &$ \pi \times 10^{-8}$    & 1 & $2\times10^{-2}$ &460 &$ 2.7\times10^{-11}$     & 3500                &  35 &  \\
D1E    &$ \pi \times 10^{-8}$    & 1 & $2\times10^{-3}$ &670 &$ \simgreat 1.0\times10^{-10}$     & 3500                &  40 &  \\
D1F    &$ \pi \times 10^{-8}$    & 1 & $2\times10^{-4}$ &420 &$ \simgreat 4.8\times10^{-10}$     & 3500                &  60 &  \\
D1G    &$ \pi \times 10^{-8}$    & 1 & $2\times10^{-6}$ &120 &$ \simgreat 1.9\times10^{-09}$     & 3500                &  75 &  \\
 & & & & &  & \\
E1F    &$ \pi \times 10^{-7}$    & 1 & $2\times10^{-4}$ &260 &$1.6\times10^{-9}$     & 7000                & 32   &  \\
\hline
\end{tabular}

$\ast$ We use the following convention to label our runs:
the first character in the name refers to $\MDOT_a$, (i.e., 
A, B, C, D, and E are for 0, $\pi\times 10^{-9}$, $10^{-8}$, 
$\pi\times 10^{-8}$, $\pi\times 10^{-7}$, respectively).
The second character refers to $x$ 
(i.e., 0 and 1, are for 0 and 1, respectively)
and finally the third character refers to $\beta'_w$
(i.e., A, B, C, D, E, F, and G are for $\infty$, $2\times10^0$,
$2\times10^{-1}$, $2\times10^{-2}$, $2\times10^{-3}$, $2\times10^{-4}$,
and $2\times10^{-6}$, respectively).
\end{center}

\end{document}